\documentclass[12pt]{article}
\usepackage{authblk}
\usepackage{pdfpages}
\usepackage{geometry}  
\usepackage{chngcntr}
\usepackage{mathrsfs} 
\counterwithin*{equation}{section}  
\usepackage[title,titletoc,toc]{appendix}
\usepackage[pagebackref=true]{hyperref}  
\usepackage{soul}
\usepackage{color}
\usepackage[most]{tcolorbox}
\usepackage[compat=1.1.0]{tikz-feynman}
\usepackage{tikz}
\usetikzlibrary{shapes.misc}

\tikzset{cross/.style={cross out, draw=black, minimum size=2*(#1-\pgflinewidth), inner sep=0pt, outer sep=0pt},
	cross/.default={1pt}}

 \tcbset{colback=yellow!10!white, colframe=red!50!black, 
        highlight math style= {enhanced, 
            colframe=red,colback=red!10!white,boxsep=0pt}
        }        
\usepackage{graphicx}
\usepackage{amssymb}
\usepackage{amsmath}
\usepackage{amsfonts}
\usepackage{epstopdf}

\newcommand{\p}{\partial}
\newcommand{\Dt}{\frac{D}{2}}
\newcommand{\dd}{\delta}
\newcommand{\ddp[1]}{\frac{\dd}{\dd #1}}
\newcommand{\lm}{\Lambda}

\newcommand{\lo}{\Lambda_0}
\newcommand{\hf}{\frac{1}{2}}
\newcommand{\be}{\begin{equation}}
\newcommand{\br}{\begin{eqnarray}}
\newcommand{\er}{\end{eqnarray}}
\newcommand{\ee}{\end{equation}}
\newcommand{\bt}{\begin{tabular}}
\newcommand{\et}{\end{tabular}}
\newcommand{\la}{\lambda}
\newcommand{\si}{\sigma}

\newcommand{\eps}{\epsilon}
\newcommand{\CD}{{\cal D}}
\newcommand{\DD}{\Delta}
\newcommand{\Dp}{\frac{d^Dp}{(2\pi)^D}}

\newcommand{\insN}{\frac1{\sqrt {N}}}
\newcommand{\ddt}{\frac{\mathrm d}{\mathrm d t}}
\newcommand{\diff}{\mathrm d}
\renewcommand{\theequation}{\arabic{section}.\arabic{equation}}

\title{Holographic RG from ERG: Locality and General Coordinate Invariance in the Bulk}

\author[1, 2]{Pavan Dharanipragada \thanks{\href{mailto:pavand@imsc.res.in}{pavand@imsc.res.in}}}
\author[1,2,3]{B. Sathiapalan \thanks{\href{mailto:bala@imsc.res.in}{bala@imsc.res.in}}}

\affil[1]{The Institute of Mathematical Sciences, CIT Campus, Tharamani, Chennai 600113, India}
\affil[2]{Homi Bhabha National Institute\\Training School Complex, Anushakti Nagar, Mumbai 400085, India}
\affil[3]{Chennai Mathematical Institute, H1, SIPCOT IT Park\\ Siruseri
Kelambakkam 603103,
India}

\begin{document}
\maketitle
\begin{abstract}
In earlier papers it was shown that the correct kinetic term for scalar, vector gauge field and the spin two field in $AdS_{D+1}$ space is obtained starting from the ERG equation for a $CFT_D$ perturbed by scalar composite, conserved vector current and conserved traceless energy momentum tensor respectively. In this paper interactions are studied and it is shown that a flipped version of Polchinski ERG equation that evolves towards the UV can be written down and is useful for making contact with the usual AdS/CFT prescriptions for correlation function calculations.  The scalar-scalar-spin-2  interaction in the bulk is derived from the ERG equation in the large $N$ semiclasical approximation.
  It is also shown that after mapping to AdS the interaction is local on a scale of the bare cutoff rather than the moving cutoff (which would have corresponded to the AdS scale).  The map to $AdS_{D+1}$ plays a crucial role in this locality. The local nature of the coupling  ensures that this interaction term in the bulk action is  obtained by gauge fixing a  general coordinate invariant scalar kinetic term in the bulk action. A wave function renormalization of the scalar field is found to be  required for a mutually consistent map of the two fields to $AdS_{D+1}$. 
\end{abstract}

\newpage
\tableofcontents
\newpage

\section{Introduction and Outline}
A precise realization of holography \cite{tHooft:1993,Susskind:1994} is the AdS/CFT correspondence [\cite{Maldacena}-\cite{Penedones2016}].One of the most interesting ideas that have come out of the AdS/CFT correspondence is that of Holographic RG [\cite{Akhmedov}-\cite{Mukhopadhyay:2016fre}].  It has been shown in a series of papers by one of the authors of this paper \cite{Sathiapalan:2017,Sathiapalan:2019,Sathiapalan:2020, Dharanipragada:2022} that the bulk AdS dual can be derived from first principles starting from a Polchinski ERG equation [\cite{Wilson}-\cite{Rosten:2010}] of the boundary CFT\footnote{To avoid confusion, one should emphasize that a CFT is a field theory at a fixed point of the RG and as such has no RG flow. The RG flow is for the CFT perturbed by the addition of some terms, for instance of the form $\int \la O$ where $O$ is some operator in the CFT.}.

In this paper we build on the work done in \cite{Sathiapalan:2017,Sathiapalan:2019,Sathiapalan:2020, Dharanipragada:2022}. Our goal is to construct, from ``first principles" the bulk AdS dual of the O(N) vector model in $D=3$. The O(N) vector model  has two fixed points---the Gaussian point and the Wilson-Fisher fixed point. Both these fixed point theories are conjectured to have as bulk AdS dual some variant of the higher spin theory \cite{Klebanov:2002} of the type studied by Vasiliev \cite{Vasiliev:1999ba}.

By the phrase ``from first principles" we mean the following:

We start with the boundary field theory with $D=3$. We then write down an Exact Renormalisation Group (ERG) for the evolution of single trace perturbations. These are functional differential equations. We then write down an evolution operator for this ERG equation in the form of a functional integral over a field theory in one higher dimension i.e. $D+1$ dimensions. This field theory is {\bf non-local} and has non-local kinetic and interaction terms. Because this bulk action implements an RG evolution  they are guaranteed to reproduce the correct correlators of the boundary theory.  We map this non local action by a field redefinition to an action in AdS space where the kinetic term has the standard {\bf local} form. This bulk action is by construction the AdS dual one is looking for.

We also get a well defined mathematical expression for the interaction terms. This was done for the three scalar interaction in \cite{Sathiapalan:2020}. In this paper we obtain the graviton-scalar-scalar interaction. There is no guess work involved in this procedure: One does not have to start with a general ansatz and fix the coefficients so that the boundary correlators are reproduced using the standard AdS/CFT framework. It is in this sense that we use the phrase ``from first principles". 

The result of the calculation is that the interaction terms are almost local. It turns out that with some minor modifications they can be made exactly local. This is one of the important results of this paper. It is not clear whether this procedure can be consistently extended for quartic and higher interactions or indeed for all other  cubic interactions.

We give an outline of the paper below.

{\bf Outline:}

To start with, the fixed point theory is a conformal field theory. We consider the Gaussian fixed point of the O(N) model, free field theory with the action
\begin{equation}
S=\hf \sqrt N\int \frac{\diff^3p}{(2\pi)^3}  \phi^I(p) \frac{1}{\DD_l}\phi^I(-p),
\end{equation}
where $\DD_l$ is the regulated propagator.

The free O(N) model has an infinite number of conserved currents that  are O(N) singlets, one for each even spin. They could be scalars or in the form of higher spin currents. These currents are the objects that have bulk duals. Hence their correlators are of interest. In this paper we consider two---the scalar composite $\sigma(x)=(\phi^I\phi^I)(x)$ and the energy momentum tensor composite
$\sigma_{\mu\nu}(x)=T_{\mu\nu}[\phi^I(x)]$. For obtaining their correlators, one adds pertubations of the form $\int j\si$ or $\int h_{\mu\nu}\si^{\mu\nu}$ to the free action.  
The theory then flows away from the fixed point and undergoes a non trivial RG evolution.  The correlators of these objects, even in the free theory, are nonzero, and further, require renormalisation. Thus, they have a nontrivial RG evolution. This evolution  is captured by the ERG equation. The AdS bulk action can be determined once the form of this ERG equation is obtained.

Polchinski's ERG equation that describes this flow is:
\begin{equation}
\frac{\p \Psi}{\p t} = \frac1{2\sqrt N}\int_p \dot \DD_l \frac{\dd^2}{\dd \phi^I(p)\dd\phi^I(-p)}\Psi.
\end{equation}
where $\Psi = e^{-S_I[\phi^I]}$. Here $S_I$ is the perturbation; $t$ is the RG parameter, and dot denotes derivative in $t$. Note that the form of the equation {\em does not} depend on $S_I$. It {\em does} depend on the free theory kinetic term through $\DD_l$.

\begin{figure}[h]
	\centering
	\includegraphics{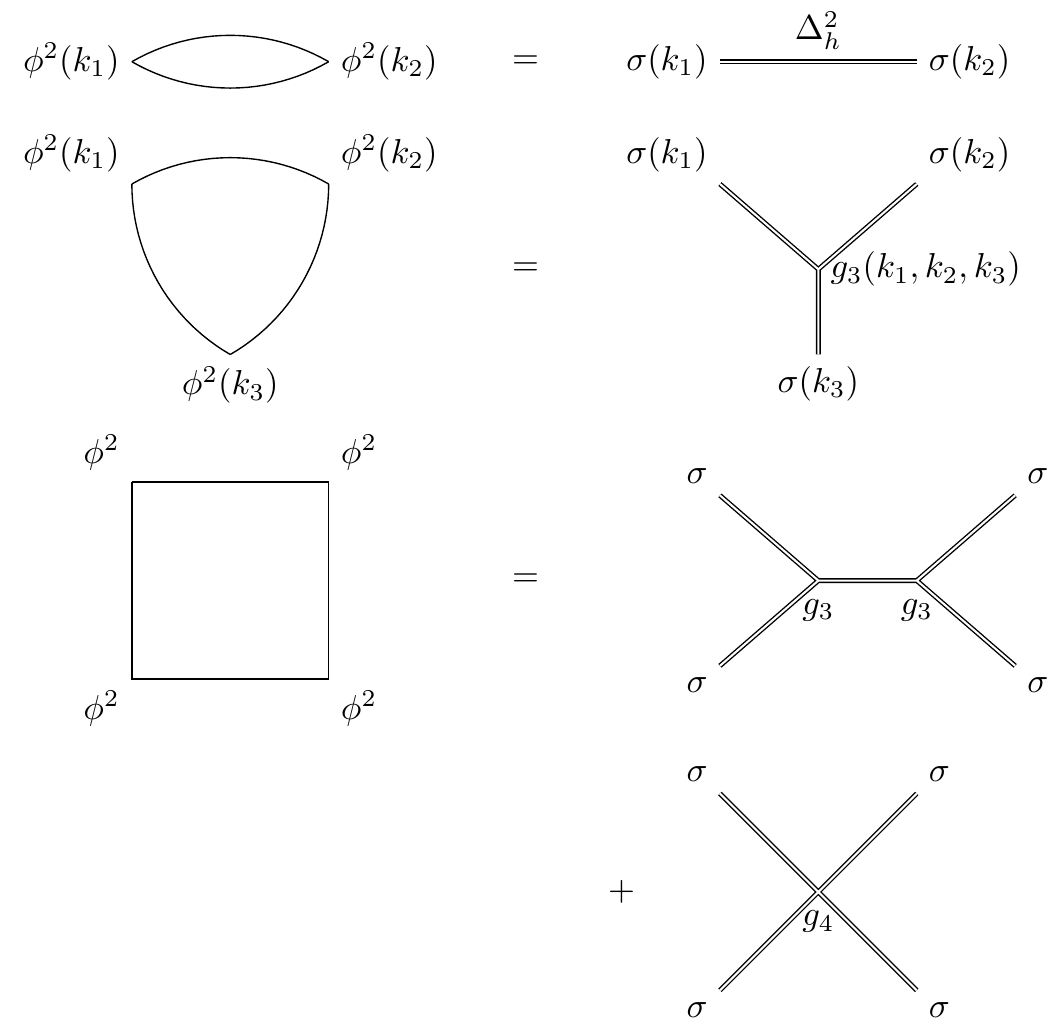}
	\caption{The vertices of $S[\sigma]$ should reproduce the correlators of $\phi^2$.}
	\label{n-pointdiagram}
\end{figure}

Since we are interested in perturbations involving $\si$ rather than $\phi^I$, it is convenient to rewrite the fixed point theory in terms of $\si$. 
Thus one  integrates out $\phi^I$ keeping $\si$ fixed and one obtains an action for $\sigma$ to be of the form
\begin{equation}
S_0[\sigma]=\frac14\int _x  \sigma(x)\frac{1}{\DD_l^2} \sigma(x) + \sum_{n=3}^\infty\int_{x_i}g_n(x_1,..,x_n,\lm)\prod_{i=3}^n\sigma(x_i).
\end{equation}
The $g_n \approx O(N^{-\frac{n}{2}+1})$ come from diagrams of the form Fig.\ref{n-pointdiagram}. The diagrams for the cconnected correlators of $\phi^2$ are shown in the LHS of the equations in Figure \ref{n-pointdiagram}. These have to be reproduced by the vertices of  $S_0[\si]$ as shown. Thus we see that $S_0[\si]$ contains interactions of all orders even though the starting theory is a free field theory for $\phi^I$.

Note that we are merely rewriting the theory at the fixed point in terms of $\si$. We have not moved away from the fixed point. To reiterate, $S_0[\sigma]$ is a nontrivial interacting fixed point action for $\sigma$, even though it comes from the free theory for $\phi^I$.

Having done this we rewrite Polchinski's equation in terms of the field $\si$. It turns out that the equation is of the form
\begin{equation}
\frac{\p \Psi}{\p t} = \Big[\int_p \frac{\diff}{\diff t}( {\DD}_l)^2 \frac{\dd^2}{\dd \si(p)\dd\si(-p)} + V[\si,\lm]\Big]\Psi,
\end{equation}
where $\Psi = e^{-S[\sigma]}$. Note that $S[\si]$ is the full action for $\si$.
This results in a non trivial flow for perturbations of the fixed point action $S_0[\si]$. Thus if we add $\int j \phi^2$ to the free field theory for $\phi^I$, this becomes the perturbation $\int j\sigma$ in $S[\si]$. Thus we can take $S[\sigma]=S_0[\si] + \int j\si$.

The potential term $V[\si,\lm]$ depends on the coefficients $g_n$ above. Once again the form of the equation depends on the fixed point action. This is expected because in AdS/CFT duality the bulk action is tied to the CFT at the boundary. 

One we have this ERG equation the evolution operator is easy to write down
\begin{equation}
\int \CD \si(t,p) \exp\Big\{-\frac14\int _t \int_p \frac{1}{2\DD_l\dot {\DD}_l} \dot \si(p,t) \dot \si(-p,t) +V[\si (p,t),t]\Big\}.
\end{equation}
Thus, the crux of the calculation is to obtain $V[\si]$. 

A field redefinition of the form $\si = f y$, along with interpreting the RG parameter $t$ as the radial coordinate, $z=e^t$, then gives us the final AdS action and thus also the interaction vertices for the AdS field $y$.\footnote{This is natural, because, in Polchinski's ERG, the parameter $t$ that appears is the logarithm of RG cutoff $\Lambda$, defined in units of the bare cutoff $\Lambda_0$, i.e., $t=\log{\Lambda/\Lambda_0}$. Thus, since the bulk radial cutoff $\epsilon$ is dual to the bare UV cutoff of the boundary theory, $z/\epsilon=\Lambda_0/\Lambda=e^t$.}
Similar steps are performed with $\si_{\mu\nu}$ and the AdS field $y_{\mu\nu}$. The main result is that (after some massaging) the cubic vertices $y^3$ and $y^2 y_{\mu\nu}$ are {\em local}. Since $y_{\mu\nu}$ is the metric tensor perturbation this is crucial for general coordinate invariance of the bulk theory.

In the (semiclassical) bulk calulation of the boundary correlation function one finds that the final  result for the correlation function is independent of the details of dependence of the Green function on the moving cutoff. This is true for any field redefinition. However for the particular field redefinition that gives AdS space it turns out that this freedom can be fruitfully used to make the interaction vertex local. This is intimately tied to the conformal nature of the correlation functions. In particular we calculate the cubic scalar-scalar-spin 2 coupling and find that it can be made local in this manner. Thus, just as with the kinetic term, the map that gives AdS space is picked out as special by the requirement of locality of the cubic coupling as well. Locality requires incorporating the correct falloff conditions for the bulk fields, i.e., the leading term in powers of $z$ for the scalar should be the boundary operator vev and for the graviton it should be the source for the operator.

The locality of the coupling of the massless spin 2 field has implications for general coordinate invariance of the theory. If one starts with a general coordinate invariant scalar kinetic term and expands about an AdS background, the linear coupling of the metric perturbation in an appropriate gauge reduces to precisely this local coupling. Hence one can conclude that this term is consistent with general coordinate invariance of the theory \cite{Kabat,Arutyunov,Arefeva,Liu,Mueck}. It also implies that if one performs a coordinate transformation, the modification of the AdS background metric can be absorbed by the dynamical graviton as a gauge transformation---equivalently it changes the fixed gauge to some other gauge.

A few technical comments are in order:
\begin{enumerate}
	\item A large $N$ expansion is essential in obtaining $V[\si]$. The ERG equation otherwise would involve higher order functional derivatives.
	\item A conceptual issue in relating Wilsonian ERG to AdS/CFT is that the AdS/CFT prescription involves integration of the bulk moving outwards to large radius---{\em towards the UV} \cite{Faulkner}. Wilsonian RG naturally starts with a UV region and moves inward. We show in this paper that a flipped ERG equation that evolves to the UV can easily be written down simply by interchanging the low and high energy propagators. They lead to the same results. This equation is unnatural physically from the point of view of RG, but is mathematically valid and corresponds more naturally to the usual AdS/CFT holographic RG calculations. In holographic RG one places the boundary at $z=\epsilon$ and imposes boundary conditions on this surface and, (in pure AdS which is what we consider in this paper), at $z=\infty$ one requires the fields to vanish. $\eps$ is taken to zero at the end. In the field theory this corresponds to taking the UV cutoff $\lm \to \infty$. We evaluate the cubic interaction using the flipped UV version of the ERG equation. While we have incorporated the cutoff $\epsilon$ in the calculations for obtaining the expressions, these must be renormalised as prescribed in \cite{Skenderis:2002wp} to get finite expressions.
	\item To see explcitly that $S_0[\si]$ is a fixed point theory it is easier to work with the generating functional $Z[J]$. This is easily shown to satisfy a fixed point equation provided we assign to $J$ a correct scaling dimension \cite{Dutta:2021yuz}.
	\item One of the technical issues one faces in calculating the graviton-scalar-scalar coupling is that of consistently mapping the kinetic terms of both the composite scalar and spin 2 composite energy momentum tensor (which becomes a graviton in the bulk) to the correct AdS form. As was pointed out in \cite{Dharanipragada:2022}, since both kinetic terms are made up of the fundamental scalar kinetic term with a specific cutoff function, it is not possible to map them both simultaneously. The way out of this problem is to perform an additional wave function renormalization of the composite scalar field. This gives some freedom to modify the cutoff dependence of the composite scalar kinetic term at high energies.  It becomes possible then to map both kinetic terms to the AdS form simultaneously.
\end{enumerate}

This work, along with \cite{Sathiapalan:2017,Sathiapalan:2019,Sathiapalan:2020, Dharanipragada:2022}, has similar goals as \cite{Aharony:2020omh}, and earlier work using ERG for building the AdS higher spin theory from the O(N) vector model \cite{Jin:2015,Douglas:2010,Leigh:2014t,Leigh:2014q}. See also \cite{SSLee:2010,SSLee:2012}. We hope to extend our formalism to all the higher spins and their cubic, quartic interactions in the future, and to understand the connection of our work with the above mentioned works.

This paper is organized as follows: In Section \ref{General-Coordinate-Invariance}, we discuss the general coordinate invariance of the theory and the issue of gauge (coordinate) transformations.  In Section \ref{uv-ir}, the flipped ERG equation  is worked out and the connection with the usual formulations of the AdS/CFT correspondence is explained.  In Section \ref{locality-scalar} the locality of the cubic scalar coupling is explained. The map to AdS space is also given and it is shown that it leads to a local interaction term in the bulk. Section \ref{graviton-coupling} discusses this issue for the graviton-scalar-scalar coupling, and section \ref{mapping-AdS} has the mapping of this term to AdS. Section \ref{summary} contains a summary, conclusions and some open questions. The appendices contain supplementary calculations, viz., the mapping to AdS for the tensor in \ref{tensormapping}, the issue of consistency of simultaneously mapping the scalar and the tensor by our procedure in \ref{field-redefinitions}, and the computation of the scalar-scalar-tensor vertex integral in \ref{appendix}.

\section{A Note on Notation}
In several places below we omit the measure in integrals to make the equations more readable. Instead we indicate the parameter to be integrated over with a subscript, like so: $\int_x$. More specifically, when a momentum is being integrated over, 
\begin{equation}
	\int_p\equiv \int\frac{\diff^D p}{(2\pi)^D},
\end{equation}
where $D$ is the number of dimensions of the boundary manifold.

\section{General Coordinate Invariance}
\label{General-Coordinate-Invariance}
\subsection{Gauge fixing a General Coordinate Invariant Action}
\subsubsection{Free Theory}

In \cite{Dharanipragada:2022}  it was shown that starting from the ERG equation for a free O(N) scalar field theory  
perturbed by the term $h^{\mu\nu}T_{\mu\nu}$, (where $T_{\mu\nu}$ is the improved energy momentum tensor), in the boundary of AdS, one can construct a bulk dual action for a free massless spin 2 field in the AdS bulk. In the boundary theory this spin 2 tensor is an auxiliary field standing for the traceless conserved energy momentum tensor composite field. In the bulk a massless spin 2 field must be a graviton. This is confirmed by comparing this action with a bulk calculation. To be more precise, one starts with the Einstein action and takes the metric in the form $(g_{AdS})_{MN} +h_{MN}$, where $(g_{AdS})_{MN}$ ($M=1,...,D+1$ )is the background $AdS_{D+1}$ metric given by
\[
ds^2= \frac{dz^2 + \dd_{\mu\nu}dx^\mu dx^\nu}{z^2}.
\]
and $h_{MN}$ is a small deviation. Extract the quadratic action for $h_{MN}$ from the Einstein action. One can set ($x^M=(z,x^\mu)$) (in the gauge $h_{zM}=0$)
\be \label{gauge} 
h_{zz}=0=h_{\mu z}= \p^\mu h_{\mu\nu}=h^\mu_\mu.
\ee  
This is a transverse-traceless gauge. The quadratic action so obtained, i.e. by gauge fixing a general coordinate invariant action is \cite{Kabat,Arutyunov,Arefeva,Liu}
\be 
S=\hf\int dz \Dp~z^{1-D}[z^4 \p _z h_{\mu\nu}\p^z h^{\mu\nu}+\p _\rho h_{\mu\nu}\p ^\rho h^{\mu\nu} + 4z^3 h_{\mu\nu} \p_z h^{\mu\nu} + 4z^2 h_{\mu\nu}h^{\mu\nu}].
\ee
Indices are raised in the above action using $\dd^{MN}$.
Setting $h_{\mu\nu}z^2 = y_{\mu\nu}$,
\be 
S=\hf\int d^{D+1}x~
z^{-D+1} ~[\p_z y_{\mu\nu}\p^z y^{\mu\nu}+\p_\rho y_{\mu\nu}\p^\rho y^{\mu\nu}].
\ee

 This is compared with what is obtained from ERG and one finds that there is agreement \cite{Dharanipragada:2022}. Thus one can say that at the free level, a dynamical gravity consistent with general coordinate invariance is obtained starting from the ERG equation of a boundary CFT in flat space.  That dynamical gravity emerges from ERG is conceptually remarkable.
 
\subsubsection{Interactions}
One can further test general coordinate invariance at the interacting level as follows. The ERG procedure applied to a scalar composite of the boundary theory generates an action \cite{Sathiapalan:2020}
\be 
S=\hf\int _x \sqrt{g_{AdS}}~ g_{AdS}^{MN}\p_M\phi \p_N \phi.
\ee
Here $\int_x\equiv\int \diff^{D+1} x$.

Let us analyse this from the bulk viewpoint. Our bulk starting point is a general coordinate invariant action:
\be 
S=\hf\int _x \sqrt{g}~ g^{MN}\p_M\phi \p_N \phi.
\ee
If we expand about $AdS$ background $g_{MN}= (g_{AdS})_{MN}+h_{MN}$ one expects a graviton-scalar-scalar coupling of the form

\be 
S=\hf\int _x \sqrt{g_{AdS}}~[ g_{AdS}^{MN}\p_M\phi \p_N \phi+~ h^{MN}\p_M\phi \p_N \phi].
\ee
(The $\sqrt g$ does not contribute a term linear in $h$ if it is traceless.) In the gauge \eqref{gauge} the interaction term becomes
\be   \label{gss}
S=\hf\int _x \sqrt{g_{AdS}}~ h^{\mu\nu}\p_\mu\phi \p_\nu \phi.
\ee
 In this paper we will verify that such a coupling does indeed exist.  This shows that the action obtained from ERG is general coordinate invariant to this order.

Since the action obtained in the bulk using the ERG procedure can be obtained by gauge fixing a general coordinate invariant theory, one concludes that the bulk theory obtained by
the ERG procedure is indeed a general coordinate invariant theory of gravity. One can then ask whether it is possible to change gauge from \eqref{gauge} to some other gauge. This should be equivalent to a coordinate transformation. Thus  
we are interested in studying the change in form due to a coordinate transformation (change of variables).

We start with
\[
S=\hf\int _x \sqrt{G}~ G^{MN}\p_M\phi \p_N \phi.
\]
Here $G_{MN}$ has the functional form of an $AdS$ metric but is not a dynamical field. Thus to begin with $G_{MN}=(g_{AdS})_{MN}$.
Under a change of coordinates 
\[
G^{MN}(x)=G^{MN}(x'+\eps(x'))= G^{MN}(x')+ \eps^P(x') \p_P G^{MN}(x'),
\]
\[
\frac{\p \phi(x)}{\p x^M}=\bigg(\dd^Q_M-\frac{\p \eps^Q}{\p x^M}\bigg)\frac{\p \phi(x'+\eps)}{\p x'^Q},
\]
\[
\sqrt G G^{MN}(x)\p_M\phi(x) \p_N \phi(x) = \sqrt GG^{MN} \frac{\p \phi(x')}{\p x'^M}\frac{\p \phi(x')}{\p x'^N}+
 \eps ^R\p_R\bigg(\sqrt G G^{MN}\frac{\p \phi(x')}{\p x'^M}\frac{\p \phi(x')}{\p x'^N}\bigg)
\]
\[ - \sqrt G\bigg(\frac{\p \eps^M}{\p x^Q}G^{QN} +\frac{\p \eps^N}{\p x^P}G^{MP}\bigg)\frac{\p \phi(x')}{\p x'^M}\frac{\p \phi(x')}{\p x'^N}.
\]
Similarly,
\[
\int d^{D+1}x = \int d^{D+1}x' (1+\p_R\eps^R).
\]
This modified action is physically equivalent to the original action because all we have done is a change of variables. But it is not manifestly invariant. Because the metric is not a dynamical variable. If we drop the boundary term, (assuming that $\eps^R$ vanishes at the boundary---small gauge transformations), then we are left with a change ($x$ derivative has been changed to $x'$ to this order in $\eps$):
\[
 -\int_{x'} \sqrt G\bigg(\frac{\p \eps^M}{\p x'^Q}G^{QN} +\frac{\p \eps^N}{\p x'^P}G^{MP}\bigg)\frac{\p \phi(x')}{\p x'^M}\frac{\p \phi(x')}{\p x'^N}.
\]
If we find a coupling to a dynamical spin two field of the form
\be  \label{y}
\dd S=\hf \int _x \sqrt G ~h^{MN}\p_M\phi \p_N\phi
\ee
then this change can be absorbed into a change of field variable
\be  \label{gty} 
h^{MN}-\frac{\p \eps^M}{\p x'^Q}G^{QN} -\frac{\p \eps^N}{\p x'^P}G^{MP}=h'^{MN}.
\ee
Since $h$ is an integration variable this doesn't affect anything and our action is now manifestly general coordinate transformation invariant---to this order in $h$. At higher orders $\sqrt G$ has to become $\sqrt {G+h}$. This is the linearized gauge transformation of the spin 2 graviton field.

Below we show that we do obtain a coupling given in \eqref{y}, but in a specific gauge where $h^{zz}=h^{\mu z}=\p_\mu h^{\mu\nu}=h^\mu_{~\mu}=0$. The gauge transformation \eqref{gty} then takes us out of this gauge. 
Thus the spin 2 field in the new gauge is:
\be 
h'^{zz}= -2z^2\frac{\p \eps^z}{\p z},~~~~ h'^{\mu z}= -z^2( \p^\mu \eps ^z + \p^z \eps^\mu),~~~~h'^{\mu\nu}= h^{\mu\nu}- z^2\p^{(\mu}\eps^{\nu )} 
\ee

\section{UV-IR Interchange in Polchinski ERG equation and AdS/CFT}
\label{uv-ir}
\subsection{Polchinski ERG Equation and UV-IR Interchange}

Let us start with the $O(N)$ vector field theory (see \cite{Zinn-Justin, Moshe:2003} for reviews):
\be	\label{1}
Z[J]=\int \CD \phi \  e^{-\hf\sqrt {N}\int_p \phi^I(p) \DD^{-1}\phi^I(-p) -S_{B,I}[\phi]+ \int J^I\phi^I }.
\ee
$S_{B,I}$ contains interaction terms in the bare action. The index $I$ runs over the first $N$ positive integers. The unmarked integrals are $D$ dimensional momentum integrals $\int \diff^D p/(2\pi)^D$.

We define $\phi_l,\phi_h$ by
\be  \label{2}
\DD = \DD_l+\DD_h,~~~\phi=\phi_l+\phi_h
\ee
with corresponding propagators $\DD_l$ and $\DD_h$. $l,h$ stand for ``low" and ``high". $\DD_l,\DD_h$ will be chosen to propagate only low/high momentum modes. We leave the propagators unspecified for the moment.

Up to some field independent factors $Z[J]$ can be written as \cite{Bagnuls1,Bagnuls2,Igarashi,Rosten:2010} :
\be  \label{3}
Z[J]= \int \CD  \phi_l \int \CD \phi_h e^{-\hf\sqrt {N}\int \phi^I_l\DD_l^{-1} \phi^I_l -\hf\sqrt {N}\int \phi^I_h\DD_h^{-1} \phi^I_h -S_{B,I}[\phi_l+\phi_h]+ \int J(\phi_l+\phi_h)}. 
\ee

The derivation of Polchinski's ERG equation proceeds {as follows: The high energy mode $\phi_h$ is integrated out and what remains is the Wilson action for $\phi_l$. Separating out the kinetic term for $\phi_l$ defines the interacting part of an action:}
\be  \label{4}
\Psi_l[\phi_l,J]\equiv e^{-S_{I,\lm}[\phi_l,J]}\equiv
\int \CD \phi_h~ e^{ -\hf\sqrt {N}\int \phi^I_h\DD_h^{-1} \phi^I_h -S_{B,I}[\phi_l+\phi_h]+ \int J(\phi_l+\phi_h)}.
\ee

Let us evaluate 
\[
\frac{\dd \Psi_l}{\dd \phi^I_l(p)}=\int \CD \phi_h 
e^{ -\hf\sqrt {N}\int \phi^J_h\DD_h^{-1} \phi^J_h}\ddp[\phi^I_l]e^{ -S_{B,I}[\phi_l+\phi_h]+ \int J(\phi_l+\phi_h)} 
\]
\[=
\int \CD \phi_h 
e^{ -\hf\sqrt {N}\int \phi^J_h\DD_h^{-1} \phi^J_h}\ddp[\phi^I_h]e^{ -S_{B,I}[\phi_l+\phi_h]+ \int J(\phi_l+\phi_h)}
\]
\[=
-\int \CD \phi_h 
\ddp[\phi^I_h(p)](e^{ -\hf\sqrt {N}\int \phi^J_h\DD_h^{-1} \phi^J_h})e^{ -S_{B,I}[\phi_l+\phi_h]+ \int J(\phi_l+\phi_h)}
\]
\[=\sqrt {N}
\int \CD \phi_h ~
\DD_h^{-1}\phi^I_h(-p)e^{ -\hf\sqrt N\int \phi^J_h\DD_h^{-1} \phi^J_h}e^{ -S_{B,I}[\phi_l+\phi_h]+ \int J(\phi_l+\phi_h)}.
\]
Then, ignoring a field independent term,
\[
\frac{\dd^2 \Psi_l}{\dd \phi^I_l(p)\dd \phi^I_l(-p)}=N
\int \CD \phi_h ~
\DD_h^{-2}\phi^I_h(-p)\phi^I_h(p)e^{ -\hf\sqrt N\int \phi^J_h\DD_h^{-1} \phi^J_h}e^{ -S_{B,I}[\phi_l+\phi_h]+ \int J(\phi_l+\phi_h)}.
\]
Now find that, (using $\frac{\p \DD_h^{-1}}{\p t}= -\frac{\dot \DD_h}{\DD_h^2}$),
\[
\frac{\p \Psi_l}{\p t}=\ \sqrt {N}
\int \CD \phi_h ~
\bigg(\int_p\hf \phi^I_h(-p)\dot \DD_h \DD_h^{-2}\phi^I_h(p)\bigg)e^{ -\hf\int \phi^J_h\DD_h^{-1} \phi^J_h}e^{ -S_{B,I}[\phi_l+\phi_h]+ \int J(\phi_l+\phi_h)}
\]
\[=
\hf \insN\int _p \dot \DD_h(p)\frac{\dd^2 \Psi_l}{\dd \phi^I_l(p)\dd \phi^I_l(-p)}.
\]

This gives {\bf Polchinski's ERG equation:}
\be   \label{perg}
\frac{\p \Psi_l}{\p t}=\hf \insN\int _p \dot \DD_h(p)\frac{\dd^2 \Psi_l}{\dd \phi^I_l(p)\dd \phi^I_l(-p)}=-\hf\insN \int _p \dot \DD_l(p)\frac{\dd^2 \Psi_l}{\dd \phi^I_l(p)\dd \phi^I_l(-p)}.
\ee
{As $\lm$ is lowered the $\phi_l$ modes get integrated over.}

Notice that mathematically there is complete symmetry between $\phi_l, \DD_l$ and $\phi_h, \DD_h$ in the way they appear in the starting action, i.e. there is a symmetry:
\be   \label{lh} 
\DD_h \leftrightarrow \DD_l~~~; ~~~\phi_h \leftrightarrow \phi_l.
\ee

{\bf Flipped Polchinski ERG Equation:}

{Thus let us repeat all the steps above but this time we integrate out $\phi_l$ first and obtain a Wilson action for $\phi_h$. We can then derive an equation describing the evolution of the Wilson action as $\lm$ is increased and the $\phi_h$ modes are integrated out. The resulting equation can be obtained simply by interchanging $l,h$ in the previous equation} \eqref{perg} {to obtain}
\be \label{pergh}
\frac{\p \Psi_h}{\p t}=\hf \insN\int _p \dot \DD_l(p)\frac{\dd^2 \Psi_h}{\dd \phi^I_h(p)\dd \phi^I_h(-p)}=-\hf \insN\int _p \dot \DD_h(p)\frac{\dd^2 \Psi_h}{\dd \phi^I_h(p)\dd \phi^I_h(-p)},
\ee
where 
\be 
\Psi_h=e^{-S_{I,\lm}[\phi_h,J]}=
\int \CD \phi_l~ e^{ -\hf\sqrt {N}\int \phi^I_l\DD_l^{-1} \phi^I_l -S_{B,I}[\phi_l+\phi_h]+ \int J(\phi_l+\phi_h)}. 
\ee

This is an action obtained by integrating out $\phi_l$ rather than $\phi_h$. We will refer to \eqref{pergh} as the flipped  ERG equation. 

{We hasten to add that this exchange given in} \eqref{lh}{ does not imply any physical symmetry between the low energy dynamics and the high energy dynamics. All that is being done is that the order of  integration has been reversed - $\phi_l$ is integrated out first and then $\phi_h$.}

So far, since $\DD_l,\DD_h$ have not been specified the physical significance of these equations have not become apparent. Let us now make a choice of propagators:
\be  \label{Delta}
\DD_l = \frac{e^{-p^2/\lm^2}}{p^2},~~~\DD_h= \frac{e^{-p^2/\lo^2}-e^{-p^2/\lm^2}}{p^2}.
\ee

Here $\lo$ (with $\lm \leq \lo$) refers to a bare cutoff and for the continuum theory one can take $\lo \to \infty$. So for our purposes
\be  \label{Delta1}
\DD_l = \frac{e^{-p^2/\lm^2}}{p^2},~~~\DD_h= \frac{1-e^{-p^2/\lm^2}}{p^2}.
\ee
Now it becomes clear that in $\Psi_l$ the high energy modes
$\phi_h$ have been integrated out and thus $S_{I,\lm}[\phi_l,J]$ is the Wilson action and \eqref{perg} is Polchinski's ERG equation. \eqref{pergh} is a mirror image of Polchinski's equation where the low energy modes have been integrated out. In \eqref{pergh}, when $\lm \to \infty$
the full functional integral has been done and we have the final answer for $Z[J]$.

Integrating $\phi_l$ may seem unphysical from a Wilsonian point of view, because one obtains non local terms, but in fact this is what a standard perturbative Feynman diagram calculation does. Consider the typical loop integral in a field theory
\[
\int _0^\lm \frac{d^4p}{(2\pi)^4} \frac{1}{p^2(p+k)^2}.
\]
It is clear here that the low energy modes with $0< |p|<\lm$ are the ones being integrated. Eventually $\lm$ is taken to infinity to get continuum results. Thus while the interpretation of a ``low energy" effective action is lost, mathematically this is a valid procedure.

In practice the low energy theory may be complicated and one may not know how to do the integral over the low energy fields due to infrared problems for instance.  One may also imagine that the low energy integral is done using the bulk theory. In any case the evolution equation continues to be valid.

\subsection{Holographic RG and  AdS/CFT}
The above  discussion has relevance to AdS/CFT calculations.

\begin{figure}[h]
	\centering
	\includegraphics[width=6cm]{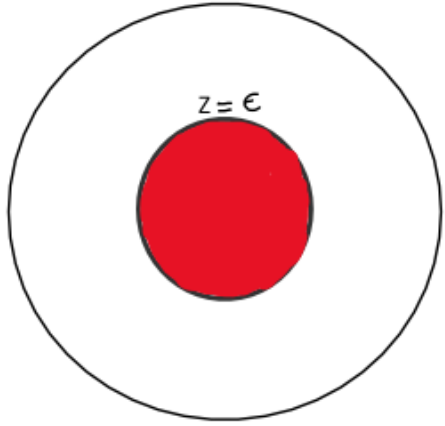}
	\caption{Integration over the red region gives $\Psi_{h,\epsilon}$. Taking $\eps	\to 0$ gives $Z[J]$. Outer boundary is at $z=0$}
	\label{Fig1}
\end{figure}

The usual AdS/CFT prescription for evaluating $Z[J]$ is depicted in Figure \ref{Fig1}. The bulk functional integral over the region $z>\epsilon$ is done with some boundary conditions at $z=\epsilon$ and this gives $\Psi_{h,\eps}$. This is equivalent to integrating out low energy modes of the dual boundary CFT. We thus set $\eps =\frac{1}{\lm}$.  The continuum limit is obtained by taking the limit $\epsilon\to 0$. The evolution of $\Psi_{h,\epsilon}$ to the outer boundary is described by \eqref{pergh}. Usual bulk holographic RG calculations also follow the situation in Figure 1.  Equations of motion are solved with boundary conditions that set the fields to zero at $z=\infty$ and boundary conditions are specified at $z=\epsilon$. The limit $\epsilon\to 0$ is taken after counterterms are added to make the action finite.

\begin{figure}[h]
	\centering
	\includegraphics[width=6cm]{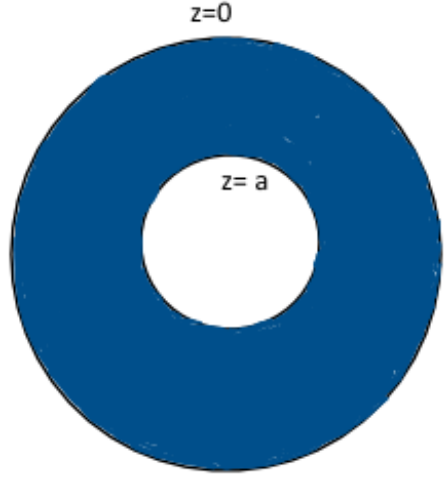}
	\caption{Integration over the blue region gives $\Psi_{l,a}$. Taking $a\to \infty$ gives $Z[J]$.}
	\label{Fig2}
\end{figure}

The Wilsonian picture  
on the other hand is \eqref{perg} and describes the situation in Figure \ref{Fig2} . The blue region has high momentum modes which are integrated over. This gives $\Psi_{l,a}$ and is the standard Wilsonian picture. Taking $a\to \infty$ gives $Z[J]$.
Both pictures are valid. For actual bulk calculations $\Psi_h$ is more useful.

We turn now to an actual calculation.

\subsection{Correlation of Scalar Composite using $\Psi_l$}

\subsubsection{ERG equation for Composite Scalar}
\label{ERGCompScalar}
We will use Polchinski ERG equation in the conventional Wilsonian form involving $\Psi_l$ first. In the next section, we repeat using $\Psi_h$.

Thus, let us start with a free $O(N)$ vector model like before:
\be	
Z[J]=\int \CD \phi  e^{-\hf\sqrt {N}\int \phi^I \DD^{-1}\phi^I +i\int J\phi^2},
\label{o(n)}
\ee
where $\phi^2$ stands for $\phi^I\phi^I$.
We have perturbed the (free) CFT by adding a source term for the composite operator $\phi^2$ \footnote{In Minkowski space there would be an i and it implements an 
	invertible functional Fourier transform. We continue with an i even in 
	Euclidean space\textemdash{}it is unconventional but legitimate. It makes it 
	invertible just as in Minkowski space. Physics is not altered by this 
	device.}.  Our strategy will be to  introduce an auxiliary field $\sigma$ to represent $\phi^2$ and determine an action $S[\sigma]$ that can be used to calculate $\sigma$-correlators using
\be
\langle \sigma(x_1)...\sigma (x_n)\rangle = \int \CD \sigma ~\sigma(x_1)...\sigma (x_n)~e^{-S[\sigma]}
\ee
The action $S[\sigma]$ can be extracted from
\[
Z[J]=\int \CD \phi \int \CD \sigma  \dd(\sigma  -\phi^2) e^{-\hf\sqrt {N}\int \phi^I \DD^{-1}\phi^I +i\int J\sigma},
\]
\be 	\label{deltainsertion}
\equiv \int \CD \sigma~ e^{-S[\sigma]+i\int J\sigma}
\ee 

We will next obtain an ERG equation for determining $S[\sigma]$.
To this end we integrate out the modes of $\phi(p)$ with $\lm < p<\infty$ as was done in Section 3.1 while deriving the Polchinski ERG equation. This will enable us to define $S_{I,\lm}[\sigma, \phi_l]$ and an ERG equation for it.

Thus we have, after introducing $\phi_l,\phi_h$ as before,
\begin{align*}
Z[J]=&\int \CD \phi_l e^{-\hf\sqrt {N}\int \phi^I_l \DD_l^{-1}\phi^I_l }\int \CD \phi_h \int \CD \sigma e^{i\int J\sigma} \int \CD \chi \exp\bigg\{i\int \chi(\sigma  -(\phi_l+\phi_h)^2)\\
&-\hf\sqrt {N}\int \phi^I_h \DD_h^{-1}\phi^I_h\bigg\}.
\end{align*}
One can do the $\phi_h$ integral:
\begin{align}
Z[J]=&\int \CD \phi_l e^{-\hf\sqrt {N}\int \phi^I_l \DD_l^{-1}\phi^I_l }\int \CD \sigma e^{i\int J\sigma} \int\CD \chi~ e^{i\int \chi\sigma }\int \CD \phi_h\nonumber\\  
&\exp\bigg\{-\hf\sqrt {N}\int \phi^I_h (\DD_h^{-1}+2i \insN\chi )\phi^I_h + 2i \int\chi   \phi^I_l\phi^I_h +i \int\chi \phi_l^2\bigg\}\nonumber\\
& =\int \CD \phi_l e^{-\hf\sqrt {N}\int \phi^I_l \DD_l^{-1}\phi^I_l }\nonumber\\
&\underbrace{\int \CD \sigma e^{i\int J\sigma}\CD \chi e^{i\int \chi\sigma }  
	\exp\bigg\{-\frac N2 Tr \ln (\DD^{-1}_h + 2i\insN\chi ) - \hf \int \phi^I_l\frac{{2}i\chi }{(1+2i\insN\DD_h\chi )} \phi^I_l\bigg\}}_{e^{-S_{I,\lm}[\phi_l,J]}\equiv \Psi_l[\phi_l,J]}.
\end{align}

Polchinski's ERG equation is, (note that $\dot \DD_h=-\dot \DD_l$):
\be \label{3PE}
\frac{\p\Psi_l}{\p t}= \hf \insN\int_p\dot \DD_h(p) \frac{\dd^2 \Psi_l}{\dd\phi^I_l(p)\dd\phi^I_l(-p)}.
\ee
Let us also define $\bar \Psi[\phi_l,\sigma]$ and $\tilde \Psi[\phi_l,\chi]$ as follows: 
\be \label{3bp}
\Psi_l[\phi_l,J] = \int \CD \sigma e^{i\int J\sigma} \bar \Psi_l[\phi_l,\sigma] = \int \CD \sigma e^{iJ\sigma} e^{-S_{I,\lm}[\phi_l,\sigma]},
\ee
so that 
\be  \label{psibar}
\bar \Psi_l[\phi_l,\sigma] =
\int\CD \chi~ e^{i\int \chi\sigma }\underbrace{
	e^{-\frac N2 Tr \ln (1+2i\insN\DD_h\chi ) - \hf \int \phi^I_l \frac{{2}i\chi}{(1+2i\insN\DD_h\chi)} \phi^I_l}}_{\tilde \Psi_l[\phi_l,\chi]}.
\ee
Thus,
\[
\Psi_l[\phi_l,J] =\int \CD \sigma e^{i\int J\sigma}  \int\CD \chi~ e^{i\int \chi\sigma }\tilde \Psi_l [\phi_l,\chi]
\]
Now we can insert this in \eqref{3PE}.  Thus, we can write 
\be \label{3PE1}
\frac{\p \tilde \Psi_l}{\p t}= \hf\insN \int_p\dot \DD_h(p) \frac{\dd^2 \tilde \Psi_l}{\dd\phi^I_l(p)\dd\phi^I_l(-p)} 
\ee
Following \cite{Sathiapalan:2020}, we will evaluate this equation at $\phi_l=0$. This simplifies things dramatically. By setting $\phi_l=0$ in the equation \emph{after} taking the derivative, we no longer track the RG flow of any interactions involving the low energy fundamental fields. If we are only interested in the RG flow of the interactions of composite operators, we can safely throw away the $\phi_l$ terms. The disadvantage is that we cannot integrate out any more $\phi_l$ modes. This means $\lm$ is a physical IR cutoff in the problem.
Thus to recover the physics, one has to set $\lm=0$. Once we have a general solution valid for arbitrary $\lm$ this is not difficult to do.

This is
\[
\hf \insN\int_p\dot \DD_h(p) \frac{\dd^2 \tilde \Psi_l}{\dd\phi^I_l(p)\dd\phi^I_l(-p)}\Bigg|_{\phi_l=0}=
- \hf\sqrt{N}\int_p \dot \DD_h(p) \bigg[\frac{{2}i\chi}{1+2i\insN\DD_h\chi}\bigg](p)\tilde \Psi_l
\]

Let us expand this in powers of $\chi$---ignoring the linear term, which comes from a tadpole diagram and can be gotten rid of by shifting the field $\sigma$ \cite{Dharanipragada:2022}. The quadratic and cubic terms are, acting on $\tilde \Psi_l$:
\[
\Big(-{2}\int_k \int _p \dot \DD_h(p)\DD_h(k+p)\chi(k)\chi(-k) + \]\[{\frac{4i}{3}}\insN\int\limits_{k_1,k_2,k_3}\dd(k_1+k_2+k_3) \int_p \frac{d}{dt}(\DD_h(p)\DD_h(k_1+p)\DD_h(k_1+k_2+p))\chi(k_1)\chi(k_2)\chi(k_3) \Big)\tilde \Psi_l.
\]
We can write the quadratic term more suggestively as
\be \label{3q}
-\int _k \frac{d}{dt} {(\DD_h^2(k))}\chi(k)\chi(-k)\tilde \Psi_l.
\ee
We get
\[
-\int_k \chi(k) \chi(-k)\frac{d}{dt}(  \DD_h^2(k))\tilde \Psi_l
+ 
\]
\be \label{3q1} 
\frac{4i}{3}\insN\int\limits_{k_1,k_2,k_3}\dd(k_1+k_2+k_3)\int_p\frac{d}{dt}( \DD_h(p)\DD_h(p+k_1)\DD_h(p+k_1+k_2))\chi(k_1)\chi(k_2)\chi(k_3) \tilde\Psi_l\Big|_{\phi_l=0}.
\ee
This pattern generalizes to higher orders.
The $n$th terms are of the form, (a factor
$\int_{k_1,k_2,...k_n}\dd(k_1+...+k_n)$ is understood in each term)
\[
i^nN^{1-\frac{n}{2}}\int _p \dot \DD_h(p)\DD_h(p+k_1)...\DD_h(p+k_1+...k_{n-1})\chi(k_1)....\chi(k_n)
\]
\[=
\frac{2^{n-2}i^n}{n}N^{1-\frac{n}{2}}\int _p \frac{d}{dt}\Big( \DD_h(p)\DD_h(p+k_1)...\DD_h(p+k_1+...k_{n-1})\Big))\chi(k_1)....\chi(k_n).
\]
We have used symmetry of the integrand to introduce a factor $1/n$ in the second expression. We thus see that the term multiplying the $n$ external $\chi$'s is a total derivative. This will be important.

Let us proceed now with our analysis of the quadratic and cubic terms. It will be convenient to work with $\bar \Psi[\phi_l,\sigma]$ (defined in \eqref{psibar}) in which
we can replace $\chi$ by $-i\frac{\dd}{\dd\sigma}$. Converting $\chi \to -i\frac{\dd}{\dd\sigma}$ we get:
\[
\frac{\p \bar \Psi_l[\phi_l,\sigma]}{\p t}\Bigg|_{\phi_l=0}=\hf\int_k \frac{d}{dt}(  2\DD_h^2(k))\frac{\dd^2}{\dd\sigma(k)\dd\sigma(-k)}\bar\Psi_l[\phi_l,\sigma]|_{\phi_l=0}-
\]
\be \label{3PE2}
\frac{4}{3\sqrt N}\int\limits_{k_1,k_2,k_3}\dd(k_1+k_2+k_3)\frac{d}{dt}\Big( \int_p\DD_h(p)\DD_h(p+k_1)\DD_h(p+k_1+k_2)\Big)\frac{\dd^3\bar\Psi_l[\phi_l,\sigma]}{\dd\sigma(k_1)\dd\sigma(k_2)\dd\sigma(k_3)} \Bigg|_{\phi_l=0}.
\ee 
We see that the propagator for the composite $\sigma$ is $2\DD_h^2$ as expected.    

Let us pause and  check whether the leading order term in $\bar \Psi$ satisfies this equation. From \eqref{3bp}, we have 
\be  \label{3sl} 
\bar \Psi_l|_{\phi_l=0} = \int \CD \chi e^{i\int \chi \sigma}
e^{\int _k \chi(k)\chi(-k)  \DD_h^2(k)}
=e^{-\frac{1}{2}\int_k\frac{\sigma(k)\sigma(-k)}{2\DD_h^2(k)}}.
\ee 
We have dropped field independent terms. This clearly satisfies the equation \eqref{3PE2} to leading order. 

Acting on the leading order $\bar \Psi_l$ we can write the second term as a cubic monomial in $\sigma$:
\[
-   \frac{4}{3}\insN\int\limits_{k_1,k_2,k_3}\dd(k_1+k_2+k_3)\frac{d}{dt}\Big( \int_p\DD_h(p)\DD_h(p+k_1)\DD_h(p+k_1+k_2)\Big)
\]
\[\times\frac{\dd^3}{\dd\sigma(k_1)\dd\sigma(k_2)\dd\sigma(k_3)}\bar \Psi_l|_{\phi_l=0}
\]
\[=
\frac{4}{3}\insN\int\limits_{k_1,k_2,k_3}\dd(k_1+k_2+k_3)\frac{d}{dt}\Big( \int_p\DD_h(p)\DD_h(p+k_1)\DD_h(p+k_1+k_2)\Big)
\]
\[\times\frac18\frac{\sigma(k_1)}{\DD_h^2(k_1)}\frac{\sigma(k_2)}{\DD_h^2(k_2)}  \frac{\sigma(k_3)}{\DD_h^2(k_3)} \bar \Psi_l|_{\phi_l=0}   
\]
\be 
\equiv \insN\int\limits_{k_1,k_2,k_3}\dd(k_1+k_2+k_3)g(k_1,k_2,k_3,\lm)\sigma(k_1)\sigma(k_2)\sigma(k_3).
\ee
Thus, the ERG equation becomes to this order
\begin{align}
\frac{\p \bar \Psi_l}{\p t}\bigg|_{\phi_l=0}=&
\Big(\hf\int_k \dot G_s(k)\frac{\dd^2}{\dd\sigma(k)\dd\sigma(-k)}\nonumber\\
&+\insN\int\limits_{k_1,k_2,k_3}\dd(k_1+k_2+k_3)g(k_1,k_2,k_3,\lm)\sigma(k_1)\sigma(k_2)\sigma(k_3)\Big)\bar\Psi_l|_{\phi_l=0},
\label{3PE4}
\end{align}
where 
\[
g(k_1,k_2,k_3,\lm)={\frac{4}{3}}\frac{d}{dt}\Big( \int_p\DD_h(p)\DD_h(p+k_1)\DD_h(p+k_1+k_2)\Big)\frac{1}{G_s(k_1,t)G_s(k_2,t)G_s(k_3,t)}.
\]
We have set {$2$}$\DD_h^2=G_s$. This determines
$\bar \Psi_l |_{\phi_l=0}=e^{-S_{I,\lm}[0,\sigma]}$.
The leading order expression for $S_{I,\lm}[0,\sigma]$ is given in \eqref{3sl}.

\subsubsection{Evolution Operator}
The evolution operator for the ERG equation \eqref{3PE4} can be written as a functional integral:
\begin{align}
K[\sigma_f,t_f;\sigma_i,t_i]=&\int\limits_{\sigma(t_i)=\sigma_i}^{\sigma(t_f)=\sigma_f} \CD \sigma ~\exp\bigg\{-\hf \int_{t_i}^{t_f} dt~\bigg[\int_k\frac{\dot\sigma(k,t)\dot\sigma(-k,t)}{\dot{G}_s(k,t)}\nonumber\\
&+\insN\int\limits_{k_1,k_2,k_3}\dd(k_1+k_2+k_3)g(k_1,k_2,k_3,\lm)\sigma(k_1,t)\sigma(k_2,t)\sigma(k_3,t)\bigg]\bigg\}.
\label{K}
\end{align}
The cubic potential term has been added to the action in the path integral, with the $\sigma$ fields becoming functions of $t$.

Let us evaluate it semiclassically, order by order.
The leading EOM is
\[
\frac{d}{dt} \frac{\dot\sigma(-k,t)}{\dot G_s(k,t)}=0 \implies \dot \sigma(-k,t)= J(k)\dot G_s(k,t).
\]
Thus,
\be   \label{sig} 
\sigma(k,t)=J(k) G_s(k,t),  ~~~~(\because \sigma(k,z=0)=G_s(k,z=0)=0).
\ee
The boundary condition is required---\eqref{3sl} is ill-defined at $z=0$ unless $\sigma(k,z=0)$ vanishes. 

If we plug this solution into \eqref{3PE4}, one obtains a product of three factors of $\frac{\sigma(k,t)}{G_s(k,t)}=J(k)$, which is $t$-independent and the cubic term becomes a total derivative. At one boundary $t=\infty$, where it is non zero, we get
\be 
\insN\int_p\DD(p)\DD(p+k_1)\DD(p+k_1+k_2)J(k_1)J(k_2)J(k_3).
\ee
This is clearly the correct answer for the one loop contribution of the cubic vertex to the amplitude.

\subsection{Correlation of Scalar Composite using $\Psi_h$}  

Given the symmetry $l\leftrightarrow h$ described in \eqref{lh}, all the results of the last section can be taken over with $l,h$ interchanged.

Thus, the $\phi_l$ fields are integrated over and we get
\[
\frac{\p \bar \Psi_h}{\p t}=\hf\int_k \frac{d}{dt}(2  \DD_l^2(k))\frac{\dd^2}{\dd\sigma(k)\dd\sigma(-k)}\bar\Psi_h|_{\phi_h=0}
\]
\be \label{UVPE2}
-   \frac{4}{3}\insN\int\limits_{k_1,k_2,k_3}\dd(k_1+k_2+k_3)\frac{d}{dt}\Big( \int_p\DD_l(p)\DD_l(p+k_1)\DD_l(p+k_1+k_2)\Big)\frac{\dd^3}{\dd\sigma(k_1)\dd\sigma(k_2)\dd\sigma(k_3)} \bar\Psi_h|_{\phi_h=0}.
\ee 
\eqref{3sl} becomes:
\be  \label{sh} 
\bar \Psi_h|_{\phi_h=0} = \int \CD \chi e^{i\int \chi \sigma}
e^{\int _k \chi(k)\chi(-k)  \DD_l^2(k)}
=e^{-\frac{1}{2}\int_k\frac{\sigma(k)\sigma(-k)}{2\DD_l^2(k)}}.
\ee 

The Flipped ERG equation is:
\begin{align}
\frac{\p \bar \Psi_h}{\p t}\bigg|_{\phi_h=0}=&
\Big(\hf\int_k {\dot G}_s(k)\frac{\dd^2}{\dd\sigma(k)\dd\sigma(-k)}\nonumber\\
&+\insN\int\limits_{k_1,k_2,k_3}\dd(k_1+k_2+k_3)g(k_1,k_2,k_3,\lm)\sigma(k_1)\sigma(k_2)\sigma(k_3)\Big)\bar\Psi_h\bigg|_{\phi_h=0},
\label{UVPE4}
\end{align}
where 
\[
g(k_1,k_2,k_3,\lm)=\frac{4}{3}\frac{d}{dt}\Big( \int_p\DD_l(p)\DD_l(p+k_1)\DD_l(p+k_1+k_2)\Big)\frac{1}{G_s(k_1,t)G_s(k_2,t)G_s(k_3,t)}.
\]
We have set $2\DD_l^2=G_s$. This determines
$\bar \Psi_h |_{\phi_h=0}=e^{-S_{I,\lm}[\phi_h=0,\sigma]}$.
The leading order expression for $S_{I,\lm}[\phi_h=0,\sigma]$ is given in \eqref{sh}.

The evolution operator is the same as in \eqref{K}:
\begin{align}
K[\sigma_f,t_f;\sigma_i,t_i]=&\int\limits_{\sigma(t_i)=\sigma_i}^{\sigma(t_f)=\sigma_f} \CD \sigma ~\exp\bigg\{-\hf \int_{t_i}^{t_f} dt~\bigg[\int_k\frac{\dot\sigma(k,t)\dot\sigma(-k,t)}{\dot{G}_s(k,t)}\nonumber\\
&+\insN\int\limits_{k_1,k_2,k_3}\dd(k_1+k_2+k_3)g(k_1,k_2,k_3,\lm)\sigma(k_1,t)\sigma(k_2,t)\sigma(k_3,t)\bigg]\bigg\}.
\label{Kh}
\end{align}
Solving semi classically as before one obtains the same equations, but with the boundary condition that $\sigma$ vanishes at $\infty$ rather than zero, because $\DD_l^2$ vanishes at $\lm=0$:
\[
\frac{d}{dt} \frac{\dot\sigma(-k,t)}{\dot G_s(k,t)}=0 \implies \dot \sigma(-k,t)= J(k)\dot G_s(k,t).
\]
Thus,
\be   \label{sigh} 
\sigma(k,t)=J(k) G_s(k,t),  ~~~~(\because \sigma(k,z=\infty)=G_s(k,z=\infty)=0).
\ee
This is the boundary condition used in AdS/CFT correlation function calculations.

The calculation of the correlation function proceeds as before by plugging in this solution into the action. As before, since $\frac{\sigma(k_i,t)}{G_s(k_i,t)}=J(k_i)$ is $t$-independent,
one obtains again a total derivative, and we pick up the contribution this time, at the $t=-\infty$ ($z=0$) boundary where $\DD^2_l=\DD^2$ and the same final answer is obtained:
\be 
\insN\int_p\DD(p)\DD(p+k_1)\DD(p+k_1+k_2)J(k_1)J(k_2)J(k_3).
\ee

\section{Locality of Interaction Term and Mapping to AdS} 
\label{locality-scalar}
In the following, we work with the action resulting from the flipped ERG equation.The $D+1$ dimensional action in \eqref{Kh} has a non-standard kinetic term and following \cite{Sathiapalan:2017, Sathiapalan:2020} we will perform a field redefinition that maps it to an action in $AdS_{D+1}$ with the usual scalar kinetic term in $AdS_{D+1}$.

The redefinition \footnote{Both these fields are essentially  ``generalized free fields" as described in \cite{Duetsch}} is 
\be \label{fr}
\sigma(p,t)=f(p,t)y(p,t),
\ee with 
\be  \label{f}
\frac{1}{f}= z^\Dt (A(p)K_\nu (pz) -B(p)I_\nu(pz)),
\ee
where $\nu=|\Delta-D/2|$, $\Delta=D-2$ being the dimension of the boundary operator $\sigma(x,0)$. Near the boundary $z=\epsilon$, the leading $z$ behaviour of $y$ is
\begin{equation}
y(p,z){\sim z^{\Delta}}\sigma(p,\epsilon).
\end{equation}
Near the boundary $z\to \epsilon$, $f\to z^{\Delta}$.
Then 
\begin{equation}
A(p) = \frac{1}{\Gamma(\nu)}2^{1-\nu}p^{\nu}.
\end{equation}
Then we get
\begin{equation}
\frac 1f = \frac{1}{\Gamma(\nu)}2^{1-\nu}p^{\nu} z^\Dt K_\nu(pz) -\frac1\gamma 2^{\nu-1}\Gamma(\nu) p^{\nu}z^\Dt I_\nu(pz).
\end{equation}
The Green function $G_s$ in turn is given by
\be \label{G}
G_s(p,z)= \frac{\gamma^2p^{-2\nu} K_\nu(pz)}{\gamma K_\nu(pz)-2^{2\nu-2}\Gamma(\nu)^2I_\nu(pz)}.
\ee
It vanishes as $z\to \infty$, and
\[
G_s\to  \gamma p^{-2\nu},~~~~~~~~~pz\to 0
\]
with $\gamma$ a normalization factor.

The combination $f/G_s$ occurs in some calculations below and is given by
\be \label{fG}
\frac{f}{G_s}= {\Gamma(\nu) 2^{-1+\nu}z^{-\Dt}} \frac{p^{\nu}}{\gamma K_\nu(pz)}.
\ee

\paragraph{Kinetic Term} 
The kinetic term in \eqref{Kh} is
\be 
-\hf \int_{t_i}^{t_f} dt~\int_p~\frac{\dot\sigma(p,t)\dot\sigma(-p,t)}{\dot{G}_s(p,t)}.
\label{KineticTerm}
\ee
Making the field redefinition \eqref{fr} with $f$ given by \eqref{f} converts it to
\be 
S_0=\hf \int dz~\int_p~ z^{-D+1}[\p_zy(p,z) \p_z y(-p,z)+(p^2+\frac{m^2}{z^2})y(p,z)y(-p,z)],
\ee
with $m^2=\nu^2-\frac{D^2}{4}=4-2D$.

Now we proceed to the interaction term.
\paragraph{Interaction Term}

\be \label{int} 
S_I=\insN\int dt~\int\limits_{k_1,k_2,k_3}\dd(k_1+k_2+k_3)g(k_1,k_2,k_3)\sigma(k_1,t)\sigma(k_2,t)\sigma(k_3,t),
\ee
with
\be \label{g}
g(k_1,k_2,k_3,\lm)=\frac{4}{3}\frac{d}{dt}\Big( \int_p\DD_l(p)\DD_l(p+k_1)\DD_l(p+k_1+k_2)\Big)\frac{1}{G_s(k_1,t)G_s(k_2,t)G_s(k_3,t)}.
\ee

We can now substitute \eqref{fr} into \eqref{int}. We will also perform one simplification. 
Consider the integral
\[
I(k_1,k_2,\lm)=\int_p\DD_l(p)\DD_l(p+k_1)\DD_l(p+k_1+k_2).
\]
We have seen that when the on shell action is used to evaluate the correlation function, only the {\em boundary value} of $I$, namely when $\lm=\lo \to \infty$, enters the final answer.  Thus we have the freedom to choose any regularization procedure for evaluating $I$ as long as it gives the correct answer when $\lm=\lo\to \infty$. So we use this freedom and modify  the regulator in $g(k_1,k_2,k_3,\lm)$.  
Thus let us choose a regulator and define
\[
I_{modified}=\bigg[\int_p \DD(p)\DD(p+k_1)\DD(p+k_1+k_2)\bigg]_{regulated,\lm}.
\]
In this particular case, (near $D=3$), one can take $\lm\to \lo\to \infty$ and the result is finite.
It is clear that this differs from the ERG prescription by
$O(\frac{k_i}{\lm})$, i.e., 
\[
I_{modified}=I+O(\frac{k_i}{\lm}).
\]
Since only the value of $I$ at $\lm=\lo \to \infty$ enters the final result, the error in the correlation function is $\frac{k_i}{\lo}$ and goes to zero as $\lo\to \infty$. Thus we are free to use any $I_{modified}$.

Now the interaction term becomes
\[
\frac{4}{3}\insN\int dt~\int_{k_1,k_2,k_3}\dd(k_1+k_2+k_3)\frac{d}{dt}\Big( \Big[\int_p\DD(p)\DD(p+k_1)\DD(p+k_1+k_2)\Big]_{regulated,\lm}\Big)
\]
\be  \label{int1}
\times\frac{f(k_1,t)f(k_2,t)f(k_3,t)}{G_s(k_1,t)G_s(k_2,t)G_s(k_3,t)}\times y(k_1,t)y(k_2,t)y(k_3,t).
\ee
The integral has been evaluated in the appendix \ref{appendix} for a  convenient and commonly used form of the regulator, and one finds:
\begin{align*}
\frac{dI_{modified}}{dt}=&\frac{d}{dt}\Big[\int_p\DD(p)\DD(p+k_1)\DD(p+k_1+k_2)\Big]_{regulated,\lm}\\
=&{\cal N}
\frac{K_\nu(k_1z)}{k_1^\nu}\frac{K_\nu(k_2z)}{k_2^\nu}\frac{K_\nu(k_3z)}{k_3^\nu}z^{2(D-3)+3\nu},
\end{align*}
where $\cal N$ is some numerical constant.
The value of $f/G_s$ is given in \eqref{fG}. Plugging all this into \eqref{int1}, one finds that there is an exact cancellation of all momentum dependence! The interaction is local
\be 
S_I={\cal N'}\int dt~\int_{k_1,k_2,k_3}\dd(k_1+k_2+k_3)y(k_1,t)y(k_2,t)y(k_3,t),
\ee
with possible cutoff dependence of  order $\frac{p}{\lo}$ where $\lo \to \infty$ is the bare cutoff which is to be taken to infinity
. The precise form of $f/G_s$ in \eqref{fG} was required for this to happen. Thus the nonlocal factor coming from the conformal momentum integrals is exactly compensated for by the function $f/G_s$ only when the map is to AdS space.

This concludes our discussion of the cubic scalar vertex. The same logic and techniques will be used for the scalar-scalar-spin 2 vertex. There is a subtlety that needs to be considered before we can do field redefinitions for both scalar and the tensor simultaneously. This is discussed in detail in appendix \ref{field-redefinitions}.

\section{Graviton Coupling}
\label{graviton-coupling}
In this section we calculate the cubic correction, involving one spin-2 composite and two scalars, to the ERG equation for the scalar composites in the $O(N)$ model. We will repeat the procedure described in \cite{Sathiapalan:2020} {\it mutatis mutandis}.

\subsection{Scalar and Tensor auxiliary fields in the Bare action}

We work around the Gaussian fixed point as in \cite{Dharanipragada:2023} for simplicity. The free massless spin-2 kinetic term in AdS background was obtained there, starting from the ERG equation for  the action for the auxiliary field that stood for the energy momentum tensor composite operator. An auxiliary field standing for the scalar composite will also be introduced. This, in turn, is very similar to the calculation in \cite{Sathiapalan:2020} except that the Wilson-Fisher fixed point action was used there.

Our starting point is thus a generating function 
\be \label{Z}
Z[J,h^{\mu\nu}] = \int \CD \phi~ e^{-\hf \int \phi^I \DD ^{-1} \phi^I -S_{B,I}[\phi,J,h^{\mu\nu}]}.
\ee
$\DD$ is a UV regulated propagator of the bare theory with a cutoff $\lo$. $J$ is a source for the composite $\phi^2$ and $h^{\mu\nu}$ is a background metric that can be used to define the energy momentum tensor $T_{\mu\nu}$.
For the free theory we take \footnote{The factor $i$ is unusual. For real sources it is a Fourier transform and thus invertible.}
\be  \label{fbi}
S_{B,I} = -i\int J \phi^2 -i \int h^{\mu\nu}T_{\mu\nu}[\phi].
\ee
$T_{\mu\nu}[\phi]$ is the traceless and conserved energy momentum tensor of the free theory.
It is given by
\be \label{T}
T_{\mu\nu}[\phi]= \p _\mu \phi^I \p_\nu \phi^I -\hf \dd_{\mu\nu} \p_\alpha \phi^I \p^\alpha\phi^I -\frac{D-2}{4(D-1)} (\p_\mu\p_\nu -\dd_{\mu\nu}\Box)\phi^2,
\ee
and satisfies 
\be \label{tt}
\p^\mu T_{\mu\nu}[\phi]=T^\mu_{~\mu}[\phi]=0.
\ee
We can then restrict $h^{\mu\nu}$ by the ``gauge" choice:
\be \label{gc1}
\p_\mu h^{\mu\nu}=h^\mu_{~\mu}=0.
\ee
Once these constraints are imposed on $h^{\mu\nu}$ it follows that only the first term in the expression  \eqref{T} for the energy momentum tensor participates in all further computations.

Introduce auxiliary fields $\sigma,\sigma_{\mu\nu}$ via delta functions
\[
\int \CD \sigma \int \CD \sigma_{\mu\nu} ~\dd(\sigma-\phi^2)\dd(\si_{\mu\nu}+i\frac{\delta}{\delta h_{\mu\nu}})=1,
\]
and $\chi,\chi^{\mu\nu}$, Lagrange multiplier fields to implement the delta functions. 
Because of \eqref{tt}
\be 
\p^\mu\sigma_{\mu\nu}=\sigma^\mu_{~\mu}=0,
\ee
and because of \eqref{gc1} $\chi^{\mu\nu}$ can also be chosen to satisfy
\be 
\p_\mu\chi^{\mu\nu}=\chi^\mu_{~\mu}=0.
\ee
Then
\[
Z[J,h^{\mu\nu}]
\]
\[ = \int \CD \phi \int \CD \sigma \int \CD \chi e^{i\int \chi(\sigma-\phi^2)} \int \CD \sigma_{\mu\nu}\int \CD \chi^{\mu\nu} e^{i\int \chi^{\mu\nu}(\sigma_{\mu\nu}+i\frac{\dd}{\dd h^{\mu\nu}})}~ e^{-\hf \sqrt N\int \phi^I \DD ^{-1} \phi^I -S_{B,I}[\phi,J,h^{\mu\nu}]}
\]
\[=
\int \CD \phi~ e^{-\hf \sqrt N\int \phi^I \DD ^{-1} \phi^I} \int \CD \sigma \int \CD \chi
\int \CD \sigma_{\mu\nu}\int \CD \chi^{\mu\nu}
e^{i\int \chi\sigma}e^{i\int \chi^{\mu\nu}\sigma_{\mu\nu}}
e^{i\int (J-\chi)\phi^2 + i\int (h^{\mu\nu}-\chi^{\mu\nu})T_{\mu\nu}[\phi]}.\]
Redefining $\chi-J \to \chi,~~ \chi^{\mu\nu}-h^{\mu\nu}\to \chi^{\mu\nu}$,  we ge
\[ Z[J,h^{\mu\nu}]=\int \CD \sigma ~e^{i\int J\sigma}
\int \CD \sigma_{\mu\nu}~ e^{ i\int h^{\mu\nu}\sigma_{\mu\nu}}
\]
\be \label{Z1} 
\underbrace{\int \CD \chi \int \CD \chi^{\mu\nu}
	\int \CD \phi ~e^{-\hf\sqrt N \int \phi^I \DD ^{-1} \phi^I}~
	e^{i\int \chi\sigma}e^{i\int \chi^{\mu\nu}\sigma_{\mu\nu}}
	e^{-i\int \chi\phi^2 - i\int \chi^{\mu\nu}T_{\mu\nu}[\phi]}}_{e^{-S[\sigma,\sigma_{\mu\nu}]}}.
\ee
$S[\sigma,\sigma_{\mu\nu}]$ is the action for the auxiliary fields obtained after integrating out $\phi$. It can be used to evaluate correlation functions of the composite operators. For instance,
\be  \label{sss}
\int \CD \sigma \int \CD \sigma_{\mu\nu}~ \sigma_{\mu \nu}(x) \sigma_{\rho\sigma}(y)e^{-S[\sigma,\sigma_{\mu\nu}]}=\langle T_{\mu\nu}[\phi](x)T_{\mu\nu}[\phi](y)\rangle.
\ee

\subsection{Wilson Action}

We now proceed to integrate out just the high energy modes of $\phi(p), ~\lo > p>\lm$ (with $\lo \to \infty$) and thus obtain the Wilson action. Thus using standard methods \cite{Igarashi} we write
\[
\phi=\phi_l+\phi_h;~~~~~~~~~\DD=\DD_l+\DD_h.
\]
The low energy propagator, $\DD_l$ propagates modes with $0<p<\lm$ and the high energy propagator $\DD_h$ propagates modes with $\lm<p<\infty$. 
\[
Z[J,h_{\mu\nu}]= \int \CD \sigma ~e^{i\int J\sigma}
\int \CD \sigma_{\mu\nu}~ e^{ i\int h^{\mu\nu}\sigma_{\mu\nu}}
\int \CD \chi \int \CD \chi^{\mu\nu}
\]
\be  \label{Z2}
\int \CD \phi_l \int \CD \phi_h~e^{-\hf \sqrt N\int \phi^I_l \DD_l ^{-1} \phi^I_l-\hf \sqrt N\int \phi^I_h \DD_h ^{-1} \phi^I_h}~
e^{i\int \chi\sigma}e^{i\int \chi^{\mu\nu}\sigma_{\mu\nu}}
e^{-i\int \chi(\phi_l+\phi_h)^2 - i\int \chi^{\mu\nu}T_{\mu\nu}[\phi_l+\phi_h]}.
\ee

The Wilson action is a functional of $\phi_l$, (and also $J,h_{\mu\nu}$), and is obtained by integrating out $\phi_h$. Thus, let us write 
\begin{align}
e^{-S_{\lm}[\phi_l,J,h_{\mu\nu}]} =& e^{-\hf \sqrt N\int \phi^I_l \DD^{-1}\phi^I_l}
\int \CD \sigma ~e^{i\int J\sigma}
\int \CD \sigma_{\mu\nu}~ e^{ i\int h^{\mu\nu}\sigma_{\mu\nu}}
\int \CD \chi \int \CD \chi^{\mu\nu}\nonumber\\
&\int \CD \phi_h~e^{ -\hf \sqrt N\int \phi^I_h \DD_h ^{-1} \phi^I_h}~
e^{i\int \chi\sigma}e^{i\int \chi^{\mu\nu}\sigma_{\mu\nu}}
e^{-i\int \chi(\phi_l+\phi_h)^2 - i\int \chi^{\mu\nu}T_{\mu\nu}[\phi_l+\phi_h]}\nonumber\\
=&e^{-\hf \sqrt N\int \phi^I_l \DD^{-1}\phi^I_l}e^{-S_{I,\lm}[\phi_l,J,h^{\mu\nu}]}.
\end{align}
Thus,
\be 
e^{-S_{I,\lm}[\phi_l,J,h^{\mu\nu}]}=\int \CD \sigma\int \CD \sigma_{\mu\nu} e^{i\int J\sigma+i\int h^{\mu\nu}\sigma_{\mu\nu}} e^{-S_{I,\lm}[\phi_l,\sigma,\sigma_{\mu\nu}]}.
\ee

$S_{I,\lm}[\phi_l,J,h^{\mu\nu}]$ as well as
its Fourier transfrom $S_{I,\lm}[\phi_l,\sigma,\sigma_{\mu\nu}]$ obey  
Polchinski ERG equation, which describes their evolution as $\lm$ is taken to zero. Thus $S[\sigma,\sigma_{\mu\nu}]$ used in \eqref{sss} is defined as
\[
S[\sigma,\sigma_{\mu\nu}]=\lim _{\lm\to 0} S_{I,\lm}[\phi_l=0,\sigma,\sigma_{\mu\nu}].
\]

Let us focus on the $\phi_h$ integration first, treating the other fields as background fields.
Define
\[
e^{-\tilde S_{I,\lm}[\phi_l,\chi,\chi^{\mu\nu}]}=\int \CD \phi_h e^{-\hf\sqrt N\int\phi^I_h\Delta_h^{-1}\phi^I_h} e^{-\tilde S_{I,\lm}[\phi_l,\phi_h,\chi,\chi^{\mu\nu}]}
\]
\[=
\int \CD \phi_h~e^{ -\hf \sqrt N\int \phi^I_h \DD_h ^{-1} \phi^I_h}~
e^{-i\int \chi(\phi_l+\phi_h)^2 - i\int \chi^{\mu\nu}T_{\mu\nu}[\phi_l+\phi_h]}.
\]
Then
\be \label{ss}
e^{-S_{I,\lm}[\phi_l,\sigma,\sigma_{\mu\nu}]}=
\int \CD \chi \int \CD \chi^{\mu\nu} e^{i\int \chi \sigma +i\int \chi^{\mu\nu}\sigma_{\mu\nu} -\tilde S_{I,\lm}[\phi_l.\chi,\chi^{\mu\nu}]}.
\ee

\subsection{Doing the \boldmath$\phi_h$ integral}

Let us isolate the various types of terms in $\tilde S_{I,\lm}[\phi_l,\phi_h,\chi,\chi^{\mu\nu}]$:
\begin{enumerate}
	\item {\bf Quadratic in \boldmath$\phi_h$}
	\begin{align}
	\tilde S_{I,\lm,hh} =&  \int_x\int_y \Big[\hf\sqrt N\phi^I_h(x)\DD_h^{-1}(x,y)\phi^I_h(y)\nonumber\\
	&+i \phi^I_h(x)\chi(x)\dd(x-y)\phi^I_h(y) +i \phi^I_h(x) \chi^{\mu\nu}(x) \Theta_{\mu\nu}\dd(x-y)\phi^I_h(y)\Big],
	\label{hh}
	\end{align}
	where $\Theta_{\mu\nu}$ is defined as the differential operator contained in $T_{\mu\nu}$, cf. \eqref{T}:
	\begin{equation}
	T_{\mu\nu}(x) \equiv \phi^I(x)\Theta_{\mu\nu}(x)\phi^I(x).
	\end{equation}
	We write this as
	\be 
	\tilde S_{\lm,hh}=\hf \int_x \int_y \phi^I_h(x) O^{-1}(x,y)\phi^I_h(y),
	\ee
	with
	\[
	O^{-1}(x,y)= \sqrt{N}\DD_h^{-1}(x,y) + \underbrace{2i \chi(x)\dd(x-y)}_{A} +\underbrace{ 2i\chi^{\mu\nu}(x)\Theta_{\mu\nu}\dd(x-y)}_{B}.
	\]

	\item {\bf Quadratic in \boldmath$ \phi_l$}
	
	\[
	\tilde S_{I,\lm,ll}=i\int \chi \phi_l^2 + i\int \chi^{\mu\nu}\phi^I_l  \Theta_{\mu\nu}\phi^I_l
	\]
	\be \label{ll}
	= i\int_x\int_y~\phi^I_l(x)\Big[ \chi(x) \dd(x-y)  + \chi^{\mu\nu}(x)\Theta_{\mu\nu}\dd(x-y)\Big]\phi^I_l(y)
	\ee
	\[
	=\hf \int_x\int_y\phi^I_l(A+B)\phi^I_l.
	\]
	\item
	{\bf Linear in \boldmath$\phi_l,\phi_h$}
	
	\[
	\tilde S_{I,\lm,lh}=2i\int \chi\phi^I_l\phi^I_h +i\int \chi^{\mu\nu}(\phi^I_l\Theta_{\mu\nu}\phi^I_h+\phi^I_h\Theta_{\mu\nu}\phi^I_l)
	\]
	\[
	= 2i\int_y\int_x \phi^I_l(x)\Big[ \chi(x)\dd(x-y) + \chi^{\mu\nu}\Theta_{\mu\nu}\dd(x-y)\Big]\phi^I_h(y)
	\]
	\[=
	\int_x\int_y\phi^I_l (A+B)\phi^I_h.
	\]
	\be 
	\equiv  \int_y j^I(y)\phi^I_h(y).
	\ee
	We have integrated by parts and used $\p_\mu\chi^{\mu\nu}=0$.
	
	Now we can do the $\phi_h$ integral in
	\[
	\int \CD \phi_h e^{-\hf \int_x\int_y\phi^I_h O^{-1}(x,y)\phi^I_h(y) -\int _x j^I(x) \phi^I_h(x)}
	\] 
	\[
	= e^{
		\frac N2 Tr \ln O +\hf\int_x\int_y j^I(x)O(x,y)j^I(y)}.
	\]
	Thus we obtain
	\[
	\tilde S_{I,\lm}[\phi_l,\chi,\chi^{\mu\nu}]=
	i\int_x\int_y~\phi^I_l(x)\Big[ \chi(x) \dd(x-y)  + \chi^{\mu\nu}(x)\Theta_{\mu\nu}\dd(x-y)\Big]\phi^I_l(y)\]
	\be 
	- \frac N2 Tr \ln O-\hf\int_x\int_y j^I(x)O(x,y)j^I(y).
	\ee
\end{enumerate}

\subsection{ERG Equation} 

If we let $\tilde\Psi = e^{-\tilde S_{I,\lm}[\phi_l,\chi,\chi^{\mu\nu}]}$ then Polchinski's ERG equation is:
\be   \label{polch}
\frac{\p \tilde \Psi}{\p t} = -\hf \insN\int_p\dot \DD_l(p) \frac{\dd^2 \tilde\Psi}{\dd \phi^I_l(p)\dd\phi^I_l(-p)}.
\ee

Following \cite{Sathiapalan:2020} we will evaluate this equation at $\phi_l=0$. This simplifies the equation considerably. As mentioned in section \ref{ERGCompScalar},  by setting $\phi_l=0$ in the equation \emph{after} taking the derivative, we no longer track the RG flow of any interactions involving the low energy fundamental fields. If we are only interested in the RG flow of the interactions of composite operators, we can safely throw away the $\phi_l$ terms. The disadvantage is that we cannot integrate out any more $\phi_l$ modes. This means $\lm$ is a physical IR cutoff in the problem.
Thus to recover the physics, one has to set $\lm=0$. Once we have a general solution valid for arbitrary $\lm$ this is not difficult to do.

This procedure gives us an equation for $\tilde S_{I,\lm}[0,\chi,\chi^{\mu\nu}]$ and after integrating over $\chi,\chi^{\mu\nu}$ as in \eqref{ss}, an equation for $S_{I,\lm}[0,\sigma,\sigma_{\mu\nu}]$.  

Define $\Psi  = e^{-\tilde S_{I,\lm}[\phi_l,\sigma,\sigma^{\mu\nu}]}=\int \CD \chi \int \CD \chi^{\mu\nu} e^{i\int \chi \sigma +i\int \chi^{\mu\nu}\sigma_{\mu\nu}}\tilde \Psi$ as in \eqref{ss}.
Then \eqref{polch}, on setting $\phi_l=0$, immediately leads to

\be   \label{polch1}
\frac{\p  \Psi}{\p t}\bigg|_{\phi_l=0} = -\hf \insN\int_p\dot \DD_l(p) \frac{\dd^2 \Psi}{\dd \phi^I_l(p)\dd\phi^I_l(-p)}\bigg|_{\phi_l=0}.
\ee
Since $\dot \DD_l =-\dot \DD_h$ we can write an equivalent equation:
\be   \label{polch2}
\frac{\p  \Psi}{\p t}\bigg|_{\phi_l=0} = \hf \insN\int_p\dot \DD_h(p) \frac{\dd^2 \Psi}{\dd \phi^I_l(p)\dd\phi^I_l(-p)}\bigg|_{\phi_l=0}.
\ee
This is a more useful form since the propagator that appears in $\Psi$ is $\DD_h$, (see \eqref{hh}).

Let us evaluate the RHS of this equation.
\[
O^{-1}= \sqrt{N}\frac{1}{\DD_h}+A+B=\sqrt{N}\DD_h^{-1}(I+\insN\DD_h A + \insN\DD_h B),
\]
\begin{align*}
O=&\insN\frac{1}{(I+\insN\DD_h A + \insN\DD_h B)}\DD_h\\
=&\insN[1-\insN(\DD_h A +\DD_h B) +
\frac1N(\DD_h A +\DD_h B)^2+...]\DD_h,
\end{align*}
\[
\hf j^IOj^I= \hf\insN\phi^I_l(A+B)[1-\insN(\DD_h A +\DD_h B) +
\frac1N(\DD_h A +\DD_h B)^2+...]\DD_h(A+B)\phi^I_l.
\]
We have to add $\hf \phi^I_l(A+B)\phi^I_l$ from \eqref{ll}. 

\[
\hf \insN\int_p\dot \DD_h(p) \frac{\dd^2 \tilde \Psi}{\dd \phi^I_l(p)\dd\phi^I_l(-p)}|_{\phi_l=0}=\]\[
\hf \int_p \dot \DD_h \{\sqrt{N}(A+B) + (A+B)[1-\insN(\DD_h A +\DD_h B) +
\frac1N(\DD_h A +\DD_h B)^2+...]\DD_h(A+B)\}\tilde \Psi.
\]
We are interested in the terms involving two $\chi$'s and one $\chi^{\mu\nu}$
\footnote{$AB$ gives term with $\chi\chi_{\mu\nu}$ and a loop integral over $\DD_h^2 p^{\mu}p^{\nu}$. This results in a $\delta^{\mu\nu}$ that kills $\chi_{\mu\nu}$ because it is traceless.
	$ABB$ gives a term with $\chi\chi_{\mu\nu}\chi_{\rho\tau}$ and a loop integral over $\DD_h^3p^\mu p^\nu p^\rho p^\tau$. This results in a 2 gravitons + scalar term which we do not investigate in this work.
}.
This will generate the required cubic term as will become clear below.
This comes from
\[
\insN\dot \DD_h\{ A[-\DD_h B]\DD_h A + B[-\DD_h A] \DD_h A + A[-\DD_h A] \DD_h B \}.
\]
This can be written compactly in position space:
\[
4i\insN\frac{d}{dt}\int\limits_{x,y,z}\chi(x) \DD_h(x-y) \chi(y) \DD_h(y-z)\chi^{\mu\nu}(z) \frac{\p^2}{\p z^\mu \p z^\nu}\DD_h(z-x).
\] 
Acting on $\Psi=e^{-S_{I,\lm}[\sigma,\sigma^{\mu\nu},\phi_l=0]}$ this term becomes
\be  \label{ssg}
-4 \insN\int\limits_{x,y,z} \frac{d}{dt}\Big(\DD_h (x-y)\DD_h(y-z) \frac{\p^2}{\p z^\mu \p z^\nu}\DD_h(z-x)\Big)
\frac{\dd^3e^{-S_{I,\lm}[\sigma,\sigma^{\mu\nu},\phi_l=0]}}{\dd \sigma(x)\dd \sigma(y)\dd \sigma_{\mu\nu}(z)}.
\ee
The Feynman diagram corresponding to this term is given below in figure \ref{feynman}.
\begin{figure}[h]
	\centering
	\includegraphics[trim={10cm 10.5cm 10cm 10.5cm}]{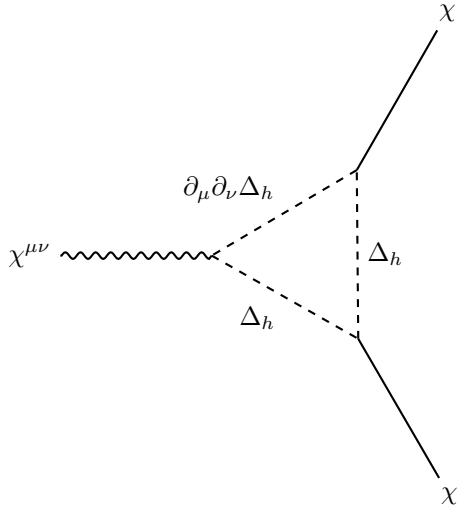}
	\caption{scalar-scalar-tensor diagram}
	\label{feynman}
\end{figure}

We now need to address a problem pointed out in \cite{Dharanipragada:2022}. {The scalar propagator
$\DD_h^2$ was fixed in {\eqref{G}} in terms of modified Bessel functions with the parameter $\nu=-\frac D2+2=\frac12$, so that the kinetic term for the scalar becomes local AdS kinetic term. This implicitly fixes the propagator $\DD_h$ for the fundamental field $\phi^I$. Similarly, for the kinetic term of the graviton to be the standard AdS kinetic term, the graviton propagator, which is proportional to
$(\p^2)^2\DD_h^2$, has to be fixed to a similar combination of modified Bessel functions with the parameter $\nu=\frac D2=3/2$, as derived in {\cite{Dharanipragada:2022}}. This would force a different constraint on $\DD_h$.} 
\begin{align}
{G_{s}(x-y,z) \equiv}& {\Delta_h(x-y)\Delta_h(x-y)=G_{1/2}(x-y,z)}\\
{G_{t} (x-y,z) =}& {\Delta_h(x-y)(\p_x^2)^2 \Delta_h(x-y)=G_{3/2}(x-y,z),}
\end{align}
{where}
\begin{equation}
{G_\nu (p,z)= \frac{\gamma p^{-2\nu} K_\nu(pz)}{ K_\nu(pz) - \frac{1}{\gamma}I_\nu(pz)}.}
\end{equation}
{These conditions are mutually incompatible.
A resolution was suggested in {\cite{Dharanipragada:2022}}. This is to redefine one of the fields (say, the scalar) by a function $g$ selected so that the propagator for the scalar is modified. This is worked out in appendix {\ref{field-redefinitions}}. Thus we let $\sigma(p)=g(p)\phi^2(p)$. Then the scalar propagator becomes $g^2\DD_h^2$. Then one can choose $g$ such that this is made up of modified Bessel functons with the parameter $\nu=-\Dt+2$\footnote{Note that in {\cite{Dharanipragada:2022}} the expression defining $g$ is slightly different. That expression is correct only when $g(p)$ is independent of $t$. In our case $g(p)$ has to depend on $t$. For $t\to-\infty$, at the boundary, $g\to1$. Thus, at boundary, $\sigma$ accurately represents our current $\phi^2$.}.} With this modification 
\eqref{ssg} changes to, (in momentum space now),
\[
-4\insN \int\limits_{k_1,k_2,k_3}\dd(k_1+k_2+k_3) \frac{d}{dt}\Big(\int_p(\DD_h (p+k_1+k_2)\DD_h(p+k_1) p^\mu p^\nu\DD_h(p)g(k_1)g(k_2))\Big)
\]
\be  \label{ssfg}
\times \frac{\dd^3}{\dd \sigma(k_1)\dd \sigma(k_2)\dd \sigma_{\mu\nu}(k_3)}e^{-S_{I,\lm}[\sigma,\sigma^{\mu\nu},\phi_l=0]}.
\ee

\subsection{ERG Equation}

In the previous sections we had introduced
the auxiliary field $\sigma$ standing for $g\phi^2$ and $\sigma_{\mu\nu}$ standing for the energy momentum tensor.
The ERG equation for the effective action involving $\sigma$  and $\sigma_{\mu\nu}$ is

\[
\frac{\p\bar \Psi_l}{\p t}\bigg|_{\phi_l=0}= \bigg\{\hf\int_k \frac{d}{dt}( g(k)^2 \DD_h^2(k))\frac{\dd^2}{\dd\sigma(k)\dd\sigma(-k)} +\hf\int_k \frac{d}{dt}( (k^2 \DD_h)^2(k))\frac{\dd^2}{\dd\sigma_{\mu\nu}(k)\dd\sigma^{\mu\nu}(-k)}
\]
\[
-   \frac{4}{3}\insN\int\limits_{k_1,k_2,k_3}\frac{d}{dt}\Big( g(k_1)g(k_2)g(k_3)\int_p\DD_h(p)\DD_h(p+k_1)\DD_h(p+k_1+k_2)\Big)\frac{\dd^3}{\dd\sigma(k_1)\dd\sigma(k_2)\dd\sigma(k_3)}
\]
\[  
-\frac{4}{\sqrt{N}} \int\limits_{k_1,k_2,k_3} \frac{d}{dt}\Big(\int_p(\DD_h (p+k_1+k_2)\DD_h(p+k_1) p^\mu p^\nu\DD_h(p)g(k_1)g(k_2))\Big)\frac{\dd^3}{\dd \sigma(k_1)\dd \sigma(k_2)\dd \sigma_{\mu\nu}(k_3)}\bigg\}
\]
\begin{equation}
\label{erg2}
\bar\Psi_l|_{\phi_l=0},
\end{equation}
where a momentum conserving $\dd(k_1+k_2+k_3)$ is implicit in the cubic terms.

The leading order solution is
\[
\bar \Psi_l|_{\phi_l=0} = \exp\bigg\{-\frac{1}{2}\int_k \frac{\sigma(k)\sigma(-k)}{G_s(k)} -
\frac{1}{2}\int_k \frac{\sigma_{\mu\nu}(k)\sigma^{\mu\nu}(-k)}{G_t(k)}\bigg\},
\]
where as before $G_s= g^2\DD_h^2$ and $G_t=(k^2\DD_h)^2$.
Acting on this, the cubic derivatives give
\[
- \frac{\sigma(k_1)}{G_s(k_1)} \frac{\sigma(k_2)}{G_s(k_2,t)} \frac{\sigma(k_3)}{G_s(k_3,t)},
\]
and
\[
- \frac{\sigma(k_1)}{G_s(k_1,t)} \frac{\sigma(k_2)}{G_s(k_2,t)} \frac{\sigma_{\mu\nu}(k_3)}{G_t(k_3,t)}.
\]
These terms add potential terms to the ERG equation.

\subsection{Flipped ERG equation}

As explained in Section 2 and Section 3 the flipped ERG equation \eqref{pergh} is more suited for mapping to AdS.
Flipping is easily done---interhange $l\leftrightarrow h$.

Thus, the flipped ERG equation is
\[
\frac{\p\bar \Psi_h}{\p t}\bigg|_{\phi_h=0}= \bigg\{\hf\int_k \frac{d}{dt}( g(k)^2 \DD_l^2(k))\frac{\dd^2}{\dd\sigma(k)\dd\sigma(-k)} + \hf\int_k \frac{d}{dt}( (k^2 \DD_l)^2(k))\frac{\dd^2}{\dd\sigma_{\mu\nu}(k)\dd\sigma^{\mu\nu}(-k)}
\]
\[
-   \frac{4}{3}\insN\int\limits_{k_1,k_2,k_3}\frac{d}{dt}\Big( g(k_1)g(k_2)g(k_3)\int_p\DD_l(p)\DD_l(p+k_1)\DD_l(p+k_1+k_2)\Big)\frac{\dd^3}{\dd\sigma(k_1)\dd\sigma(k_2)\dd\sigma(k_3)}
\]
\[
-4\insN \int\limits_{k_1,k_2,k_3} \frac{d}{dt}\Big(\int_p(\DD_l (p+k_1+k_2)\DD_l(p+k_1) p^\mu p^\nu\DD_l(p)g(k_1)g(k_2))\Big)
\]
\be 
\times\frac{\dd^3}{\dd \sigma(k_1)\dd \sigma(k_2)\dd \sigma_{\mu\nu}(k_3)}\bigg\}\bar\Psi_h|_{\phi_h=0},
\ee
with a momentum conserving $\dd(k_1+k_2+k_3)$ in the cubic terms.

The leading order solution is
\begin{equation}
\bar \Psi_h|_{\phi_h=0} = \exp\bigg\{-\frac{1}{2}\int_k \frac{\sigma(k)\sigma(-k)}{G_s(k)} -
\frac{1}{2}\int_k \frac{\sigma_{\mu\nu}(k)\sigma^{\mu\nu}(-k)}{G_t(k)}\bigg\},
\label{flippedWilsonAction}
\end{equation}
where now $G_s= g^2\DD_l^2$ and $G_t=(k^2\DD_l)^2$.
Acting on this the cubic derivative terms give
\[
- \frac{\sigma(k_1)}{G_s(k_1)} \frac{\sigma(k_2)}{G_s(k_2,t)} \frac{\sigma(k_3)}{G_s(k_3,t)},
\]
and
\[
- \frac{\sigma(k_1)}{G_s(k_1,t)} \frac{\sigma(k_2)}{G_s(k_2,t)} \frac{\sigma_{\mu\nu}(k_3)}{G_t(k_3,t)}.
\]
These contribute potential terms to the flipped ERG equation.

~
~
~
~

\section{Mapping Evolution Operator to AdS }
\label{mapping-AdS}
The evolution operator for the flipped ERG equation can be written as
\[
\int \CD \sigma \CD \sigma_{\mu\nu}~
\exp\bigg\{-\int dt\int_p~\bigg[\hf\frac{\dot \sigma \dot \sigma}{\dot G_s} +\hf\frac{\dot \sigma_{\mu\nu} \dot \sigma^{\mu\nu}}{\dot G_t} + ~ V(\sigma, \sigma_{\mu\nu})\bigg]\bigg\}
\]
where 
\[ V(\sigma, \sigma_{\mu\nu})=\insN\int\limits_{k_1,k_2,k_3}\dd(k_1+k_2+k_3)\times
\]
\[
\bigg\{-\frac{4}{3}\frac{d}{dt}\Big( g(k_1)g(k_2)g(k_3)\int_p\DD_l(p)\DD_l(p+k_1)\DD_l(p+k_1+k_2)\Big)\Big( \frac{\sigma(k_1,t)}{G_s(k_1,t)} \frac{\sigma(k_2,t)}{G_s(k_2,t)} \frac{\sigma(k_3,t)}{G_s(k_3,t)}\Big)
\]
\be \label{v}
-4  \frac{d}{dt}\Big(\int_p(\DD_l (p+k_1+k_2)\DD_l(p+k_1) p^\mu p^\nu\DD_l(p)g(k_1)g(k_2))\Big)\Big( \frac{\sigma(k_1,t)}{G_s(k_1,t)} \frac{\sigma(k_2,t)}{G_s(k_2,t)} \frac{\sigma_{\mu\nu}(k_3,t)}{G_t(k_3,t)}\Big)\bigg\}.
\ee
The scalar part of this was done in Section 3, (without the factor $g$). As shown in Section 3.1, substituting the leading order solution, one sees that the factors $\frac{\sigma(k,t)}{G_s(k,t)}$ are all time independent. Then the integrand becomes a total derivative and we recover the expected amplitudes. The important point is that the final amplitude thus depends only on the value of $g\DD_l$ at the limits of the $t$ integration---where it is zero at one end, ($z=\infty$), and $\DD$ at the other $(z=0)$. As explained in Section 3 we can modify the regularization procedure in each of the loop integrals. The errors in the final answer, viz, correlation function, is of $O(\frac{k_i}{\lo})$. In the limit $\lo\to \infty$ there is no error. We use this freedom exactly as it was done in Section 3. Thus $V$ becomes
\[ V(\sigma, \sigma_{\mu\nu})=\insN\int\limits_{k_1,k_2,k_3}\dd(k_1+k_2+k_3)\times
\]
\[
\bigg\{-\frac{4}{3}\frac{d}{dt} \Big[\int_p\DD(p)\DD(p+k_1)\DD(p+k_1+k_2)\Big]_{regulated,\lm}\Big( \frac{\sigma(k_1,t)}{G_s(k_1,t)} \frac{\sigma(k_2,t)}{G_s(k_2,t)} \frac{\sigma(k_3,t)}{G_s(k_3,t)}\Big)
\]
\be \label{v1}
-4  \frac{d}{dt}\Big[\int_p(\DD (p+k_1+k_2)\DD(p+k_1) p^\mu p^\nu\DD(p))\Big]_{regulated,\lm}\Big( \frac{\sigma(k_1,t)}{G_s(k_1,t)} \frac{\sigma(k_2,t)}{G_s(k_2,t)} \frac{\sigma_{\mu\nu}(k_3,t)}{G_t(k_3,t)}\Big)\bigg\}.
\ee

We now map the action to AdS space by the substitution  $\sigma = fy$ and $\sigma_{\mu\nu}=fy_{\mu\nu}$. Thus the coefficient of $y(k_1)y(k_2)y(k_3)$ becomes
\[
-\frac{4}{3}\frac{d}{dt} \Big[\int_p\Delta(p)\Delta(p+k_1)\Delta(p+k_1+k_2)\Big]_{regulated,\lm}\Big( \frac{f(k_1,t)}{G_s(k_1,t)} \frac{f(k_2,t)}{G_s(k_2,t)} \frac{f(k_3,t)}{G_s(k_3,t)}\Big).
\]
As mentioned in Section 3, the time derivative of the loop momentum integral with a convenient regularization procedure is,
\[
2(k_1{ \Lambda})^{-\nu_1}K_{\nu_1}(k_1/\Lambda)\times
2({k_2}{ \Lambda})^{-\nu_2}K_{\nu_2}(k_2/\Lambda  )\times
2({k_3}{ \Lambda})^{-\nu}K_{\nu_3}(k_3 /\Lambda)
\]
upto a normalising factor. 
The factor $\frac{f}{G_s}$ was given in \cite{Sathiapalan:2020} and is the inverse of the above expression. Thus there is an exact cancellation, exactly as described in Section 3 for the scalar case, since \eqref{fG} 
\[
\frac{f(k,z)}{G_s(k,z)}=\frac{z^{-D/2}(k)^\nu}{K_\nu(k/\Lambda)}.
\]
So they cancel exactly for the three scalar case and we get a local term up to some normalization.
\be  \label{scalar}
{\cal N}\int_{k_1,k_2,k_3}\dd(k_1+k_2+k_3)y(k_1,\Lambda)y(k_2,\Lambda)y(k_3,\Lambda).
\ee

For the second loop integral, the computation is done in appendix \ref{appendix}, resulting in
\begin{equation}
z^{2D-2}\Big(\frac{k_1}z\Big)^{-\nu}K_{\nu}(k_1z)~
\Big(\frac{k_2}z\Big)^{-\nu}K_{\nu}(k_2z)~
\Big(\frac{k_3}z\Big)^{\nu'}K_{\nu'}(k_3z),
\end{equation}
where $\nu$ and $\nu'$ correspond to the scalar and the tensor respectively. For the tensor, (see appendix \ref{tensormapping}),
\begin{equation}
\frac{f_t(k,z)}{G_t(k,z)}=\frac{z^{-D/2}k^{-\nu'}}{K_{\nu'}(kz)}
\end{equation}
For the other term also we have cancellation resulting in
\be \label{scalarg}
{\cal N'}z^{D-2}\int_{k_1,k_2,k_3}\dd(k_1+k_2+k_3)k_{1\mu} y(k_1,\Lambda)k_{2\nu} y(k_2,\Lambda)y^{\mu\nu}(k_3,\Lambda).
\ee
(We have used transversality of $y_{\mu\nu}$ to replace one of the $k_1$'s by $k_2$.) Above we have identified the scale $1/\Lambda$ with the radial coordinate $z$, with which the mapping is complete. The above term is precisely the coupling of the metric perturbation to the  scalar kinetic term given in \eqref{gss}. 

As explained in Section 2 the presence of this coupling verifies general coordinate invariance of the holographic action obtained from ERG.

\section{Summary and Conclusions}
\label{summary}
The most interesting aspect of the AdS-CFT correspondence is that the bulk dual of a field theory in flat space is a theory with dynamical gravity. The approach outlined in this and earlier papers demonstrates that the functional integral describing an evolution operator of the Exact RG equation of a flat space field theory involves a field theory in a background AdS space and further that the metric fluctuations about this background  are also dynamical.   That dynamical gravity comes out of ERG is quite remarkable. This demonstration did not assume the AdS-CFT conjecture.

The calculation described in this paper also explains why AdS is special. While the functional integral description of the ERG evolution operator started off with a non standard action, a very special field redefinition was required to map this to a field theory in AdS space. What we see is that when this is done the action is local on a much smaller scale than the AdS scale. The scale of locality is set by the bare cutoff of the field theory rather than the moving cutoff. In the bulk this should correspond to the Planck scale or the string scale. This was already shown for the kinetic terms in earlier papers \cite{Sathiapalan:2017, Dharanipragada:2022}. In this paper this was demonstrated for the scalar-scalar-spin 2 coupling \eqref{scalarg}. This also happens for the cubic scalar coupling\eqref{scalar}.

 The locality of the scalar gravity coupling also ensures that this interaction term can be obtained by gauge fixing a general coordinate invariant scalar kinetic term. The spin 2 kinetic term is also obtainable by gauge fixing the kinetic term for a metric perturbation obtained from the Einstein action in an AdS background \cite{Dharanipragada:2022}. Thus both these calculations provide evidence for a general coordinate invariant theory in the bulk. If this persists to higher orders, that would provide some further insight into the AdS-CFT correspondence.
 
 In deriving this cubic interaction it turned out to be useful to work with a flipped ERG equation \eqref{pergh}, that evolved the  theory towards the UV rather than the towards the IR as in the usual Wilsonian RG. It remains to be seen whether this equation has other applications.

Another ingredient in making all this work is the necessity of performing a wave function renormalization of the scalar field standing for the composite scalar. One expects that extending the ERG approach to higher spins will involve similar wave function renomrmalizations for all the fields. 
  
Extending all this to higher spins is certainly an interesting open problem. Some aspects of RG for higher spins has been discussed in \cite{Douglas:2010,Leigh:2014t,Leigh:2014q}. It would be interesting to apply our techniques to understand the higher spin action and symmetries along those lines. That would supplement holographic reconstruction of the bulk higher spin interactions from the boundary $O(N)$ model \cite{Sleight:2016dba}. A consistent application of the prescription in this paper could be applied to obtain bulk higher spin interactions and strengthen the higher spin-$O(N)$ model duality \cite{Giombi:2009}.

All our calculations have been done for Euclidean field theories. The issue of analytically continuation to Minkowski space also needs to be investigated.

These are  pressing questions and we hope to address these soon.

\begin{appendices}
\renewcommand{\theequation}{\Alph{section}.\arabic{equation}}

\section{EM tensor mapping to AdS}
\label{tensormapping}
The mapping for the tensor is similar to the mapping for the scalar. The holographic action for the tensor is to leading order
\begin{equation}
S = -\frac12\int \diff t\int_p\frac{\dot \sigma_{\mu\nu} \dot \sigma^{\mu\nu}}{\dot G_t}.
\end{equation}
We redefine
\begin{equation}
\sigma_{\mu \nu}(p,t) = y_{\mu\nu}(p,t)f(p,t),
\label{mapping}
\end{equation}
with $f^2= -\dot G_t z^{-D}/2$, $z=e^t$ and 
\begin{equation}
\label{ftensor1}
\frac{1}{f}= z^\Dt (A(p) K_\nu(pz)+B(p) I_\nu(pz)).
\end{equation}
Here $\nu=|\Delta-D/2|$, where the dimension $\Delta=D$ for the tensor. Thus, $\nu=D/2=3/2$. With this redefinition, we get an action for a set of scalars $y_{\mu\nu}$ in AdS of mass squared $m^2=\nu^2-D^2/4=0$. We get the action
\begin{equation}
\label{sads}
S_{AdS}=\int dz \Dp z^{1-D}\Big[\frac{\p y_{\mu\nu}(p)}{\p z}\frac{\p y_{\rho\sigma}(-p)}{\p z}+ p^2y_{\mu\nu}(p)y_{\rho\si}(-p)\Big]\dd^{\mu\rho}\dd^{\nu \si}.
\end{equation}
Following the same arguments in \cite{Sathiapalan:2017} section 2.4.3,
\begin{equation}
G_t (p,z)= \frac{C(p) K_\nu(pz)+D(p)I_\nu(pz)}{A(p) (K_\nu(pz) +B(p)I_\nu(pz))},
\end{equation}
and that the two linear combinations must be linearly independent gives the condition, due to the Wronskian,
\begin{equation}
A(p)D(p)-B(p)C(p)=1.
\end{equation}
Say the boundary is at $z=\epsilon=1/\Lambda_0$. Then near the boundary the leading behaviour of the bulk field is
\begin{equation}
y_{\mu\nu}(x,z)\sim {z^{D-\Delta}}h_{\mu\nu}(x),
\label{yasymptotic}
\end{equation}
where $h_{\mu\nu}$ is the source for the boundary EM tensor $T_{\mu\nu}$. 

At this stage, a small digression is necessary to figure out what the behaviour of $f$ should be near the boundary. The equation of motion in the bulk is analogous to the scalar case given in  \eqref{sig}.
\begin{equation}
\sigma_{\mu \nu}(p,z)=G_t(p,z) J_{\mu\nu}(p),
\end{equation}
for some $J_{\mu\nu}(p)$. The Wilson action at the boundary is given by \eqref{flippedWilsonAction},
\begin{equation}
S_{\epsilon}=-\int_p \frac{\sigma_{\mu \nu}(p)\sigma^{\mu \nu}(-p)}{G_t(p)}+\int_p h_{\mu\nu}(p)\sigma^{\mu\nu}(-p).
\end{equation}
The equation of motion here is
\begin{equation}
\sigma_{\mu \nu}(p,\epsilon)= G_t(p,\epsilon)h_{\mu\nu}(p).
\end{equation}
Thus we have
\begin{equation}
\sigma_{\mu \nu}(p,z)=G_t(p,z)h_{\mu\nu}(p).
\end{equation}
At boundary, $G_t(p,z)= p^{2\nu}$, so
\begin{equation}
\lim_{z\to\epsilon}\sigma_{\mu \nu}(p,z)\approx p^{2\nu}h_{\mu\nu}(p).
\end{equation}
Then from \eqref{mapping} and \eqref{yasymptotic}, it must be that
\begin{equation}
f(p,\epsilon) \approx p^{2\nu}.
\end{equation}

And since near $z\to0$,
\begin{equation}
K_\nu\to \Gamma(\nu)2^{\nu-1}(pz)^{-\nu},\quad I_\nu\to \frac{2^{-\nu}}{\Gamma(1+\nu)}(pz)^\nu,
\end{equation}
\begin{equation}
f(p,z)\sim \frac{2^{1-\nu}}{\Gamma(\nu)A(p)}z^{-\frac D 2}(pz)^{\nu}.
\end{equation}
Therefore, we get
\begin{equation}
A(p)=\frac{1}{\Gamma(\nu)}2^{1-\nu}p^{-\nu}.
\end{equation}
$G_t$ is the low energy propagator. For $z\to \infty$, $G_t$ must vanish. But for $x\to\infty$
\begin{equation}
I_\nu(x)\sim \frac{1}{\sqrt{2\pi x}}e^x, \quad K_\nu(x)\sim \sqrt{\frac{\pi}{2x}} e^{-x}.
\end{equation}
So, $D(p)=0$, and hence $B(p)C(p)=-1$.
At the boundary $G_t$ must be equal to the full propagator $G=\gamma p^{2\nu}$, i.e., for $z\to 0$, $G_t(p,z)\to \gamma p^{2\nu}$.
\begin{equation}
\frac{C(p)}{A(p)}=\gamma p^{2\nu}.
\end{equation}
\be
C(p)=\frac{\gamma}{\Gamma(\nu)}2^{1-\nu}p^{\nu}
\ee
Since $BC=-1$,
\be 
B(p)=-\frac{\Gamma(\nu)}{\gamma}2^{\nu-1}p^{-\nu}
\ee
Then
\begin{equation}
\frac{1}{f}=\frac{1}{\Gamma(\nu)}2^{1-\nu}p^{-\nu} z^\Dt K_\nu(pz) -\frac1\gamma 2^{\nu-1}\Gamma(\nu) p^{-\nu}z^\Dt I_\nu(pz),
\end{equation}
and
\begin{equation}
G_t(p,z)= \frac{\gamma^2p^{2\nu} K_\nu(pz)}{\gamma K_\nu(pz)-2^{2\nu-2}\Gamma(\nu)^2I_\nu(pz)}.
\end{equation}
Finally,
\begin{equation}
\frac{f}{G_t}=\frac{z^{-D/2}}{C(p)K_\nu(pz)}={\frac{z^{-D/2}2^{\nu-1}\Gamma(\nu)}{\gamma p^\nu K_\nu(pz)}}
\end{equation}

\section{Field Redefinitions for Scalar Composite}
\label{field-redefinitions}
Obtaining \eqref{y} involves simultaneous mapping to AdS of the scalar and tensor fields. As mentioned in \cite{Dharanipragada:2022} this requires a field redefinition of the composites. In this section we elaborate on this field redefinition.

The scalar composite was introduced in \cite{Sathiapalan:2020} by imposing the constraint $\sigma=\phi^2$. We would like to understand the freedom of a wave function renormalization of the form $\sigma(p) \to g(p,\lm)\sigma(p)$ where $g$ depends on the moving cutoff $\lm$. 

We proceed as in \ref{ERGCompScalar}, by integrating out the high energy modes and obtaining ERG equation for $\sigma$, except now we insert $\sigma=g(p,\lm)\phi^2$. Thus we modify \eqref{deltainsertion},
\be	
Z=\int \CD \phi \int \CD \sigma \dd(\sigma  -g\phi^2) e^{-\hf\sqrt N\int \phi^I \DD^{-1}\phi^I}.
\ee
Proceeding as before, separating high and low energy modes and propagators, we obtain
\be Z=\int \CD \phi_l e^{-\hf\sqrt N\int \phi^I_l \DD_l^{-1}\phi^I_l }\underbrace{\int \CD \sigma \int\CD \chi~ e^{i\int \chi\sigma }  
	e^{-\frac N2 Tr \ln (\DD^{-1}_h + 2i\insN\chi g) - \hf \int \phi^I_l\frac{i\chi g}{(1+2i\insN\DD_h \chi g)} \phi^I_l}}_{e^{-S_{I,\lm}[\phi_l]}\equiv \Psi}.
\ee 
Polchinski's ERG equation is, (note that $\dot \DD_h=-\dot \DD_l$):
\be \label{PE}
\frac{\p\Psi}{\p t}= \hf \insN\int_p\dot \DD_h(p) \frac{\dd^2 \Psi}{\dd\phi^I_l(p)\dd\phi^I_l(-p)}
\ee
Let us also define $\bar \Psi$ by
\be \label{bp}
\Psi = \int \CD \sigma  \bar \Psi
\ee
so that 
\[\bar \Psi =
\int\CD \chi~ e^{i\int \chi\sigma }\underbrace{\int 
	e^{-\hf NTr \ln (1+ 2i\insN\DD_h\chi g) - \hf \int \phi_l \frac{i\chi g}{(1+2i\insN\DD_h\chi g)} \phi_l}}_{\tilde \Psi}.
\]
Thus
\[
\Psi =\int \CD \sigma \int\CD \chi~ e^{i\int \chi\sigma }\tilde \Psi [\phi_l,\chi g].
\]

Now we can insert this in \eqref{PE}. But the time derivative gets a contribution not only from $\DD_h$ which is captured by the RHS of Polchinski ERG equation, but it has an additional dependence due to $g$. This quantity obeys the following equation.

\be \label{PE1}
\frac{\p \tilde \Psi}{\p t}= \hf \insN\int_p\Big(\dot \DD_h(p)+\frac1g\Delta_h\dot g\Big) \frac{\dd^2 \tilde \Psi}{\dd\phi_l(p)\dd\phi_l(-p)}=\hf \insN\int_p\frac1g\ddt(\Delta_hg) \frac{\dd^2 \tilde \Psi}{\dd\phi_l(p)\dd\phi_l(-p)}.
\ee
We can evaluate this at $\phi_l=0$, and the discussion follows exactly like in the section \ref{ERGCompScalar}, with $\Delta_hg$ replacing $\Delta_h$ everywhere.
The ERG equation becomes to cubic order
\begin{align}
\frac{\p \bar \Psi}{\p t}\bigg|_{\phi_l=0}=&
\Big(\int_k \dot G_s(k)\frac{\dd^2}{\dd\sigma(k)\dd\sigma(-k)}\nonumber\\
&+\insN\int\limits_{k_1,k_2,k_3}\dd(k_1+k_2+k_3) f(k_1,k_2,k_3,\lm)\sigma(k_1)\sigma(k_2)\sigma(k_3)\Big)\bar\Psi|_{\phi_l=0},
\label{PE4}
\end{align}
where 
\begin{align*}
&f(k_1,k_2,k_3,\lm)=\\
&-\frac{4}{3}\frac{d}{dt}\Big( g(k_1)g(k_2)g(k_3)\int_p\DD_h(p)\DD_h(p+k_1)\DD_h(p+k_1+k_2)\Big)\frac{1}{G_s(k_1)G_s(k_2)G_s(k_3)}.
\end{align*}
We have set $2g^2\DD_h^2=G_s$. This determines
$\bar \Psi|_{\phi_l=0}=e^{-S_{I,\lm}[0,\sigma]}$.
The leading order expression for $S_{I,\lm}[0,\sigma]$ is
\begin{equation}
\label{sl} 
\bar \Psi|_{\phi_l=0} = \int \CD \chi e^{i\int \chi \sigma}
e^{\int _k \chi(k)\chi(-k) g(k)^2 \DD_h^2(k)}
=e^{-\frac{1}{4}\int_k\frac{\sigma(k)\sigma(-k)}{g^2\DD_h^2(k)}}.
\end{equation}
Evaluating the evolution operator for this equation semiclassically as before, one obtains the same correlator:
\be 
\insN\int_p\DD(p)\DD(p+k_1)\DD(p+k_1+k_2)J(k_1)J(k_2)J(k_3).
\ee

\section{One loop scalar-scalar-graviton graph}
\label{appendix}
\paragraph{Three Point Function}
The Feynman diagram is ($k_1+k_2+k_3=0$).
\be  \label{start}
I^{\mu\nu}(k_1,k_2,k_3,\lm)=\int \Dp \frac{p^\mu p^\nu}{(k_1+p)^{2a_3}(k_1+k_2+p)^{2a_1}(p)^{2a_2}}.
\ee
with some regulator $\lm$. The regularization scheme  we leave unspecified for now. We will use 
the tracelessness and transversality of $\chi_{\mu\nu}$($\chi^\mu_{~\mu}=0=k_3^\mu\chi_{\mu\nu}(k_3)$) to simplify results.
So  effective action at the cubic order is 
\[
\Delta S_3=\int_{k_1}\int_{k_2}\int_{k_3}\chi(k_1)\chi(k_2)\chi_{\mu\nu}(k_3)I^{\mu\nu}(k_1,k_2,k_3,\lm)(2\pi)^D\delta^D(k_1+k_2+k_3).
\]
Thus we need
\be
I=\int ds_1 ds_2 ds_3 \frac{s_1^{a_1-1}s_2^{a_2-1}s_3^{a_3-1}}{\Gamma(a_1)\Gamma(a_2)\Gamma(a_3)}\int \Dp ~p^\mu p^\nu ~e^{-(k_1+p)^2s_3}e^{-(k_1+k_2+p)^2s_1}e^{-(p)^2s_2}.
\ee
Simplify exponent. Use $k_1+k_2=-k_3$.
\begin{align*}
&-(k_1+p)^2s_3-(k_1+k_2+p)^2s_1-(p)^2s_2\\
&=- [p^2(s_1+s_2+s_3)+2k_1.p s_3 - 2k_3.ps_1 +k_1^2 s_3 +k_3^2 s_1]\\
&=
-(s_1+s_2+s_3)[(p+\frac{k_1s_3-k_3s_1}{s_1+s_2+s_3})^2-(\frac{k_1s_3-k_3s_1}{s_1+s_2+s_3})^2+\frac{k_1^2 s_3+k_3^2s_1}{s_1+s_2+s_3}].
\end{align*}
Let $p'= p+\frac{k_1s_3-k_3s_1}{s_1+s_2+s_3}$.

%
Simplify exponent further using $2k_1.k_3=(k_1+k_3)^2- k_1^2-k_3^2=k_2^2-k_1^2-k_3^2$.
\begin{align*}
&\frac{k_1^2s_3^2+k_3^2s_1^2-(k_2^2-k_1^2-k_3^2)s_1s_3 - k_1^2s_3(s_1+s_2+s_3)-k_3^2s_1(s_1+s_2+s_3)}{(s_1+s_2+s_3)}\\
&=-\frac{k_1^2s_2s_3+k_3^2s_1s_2+k_2^2s_1s_3}{(s_1+s_2+s_3)}.
\end{align*}
The exponent is
\[
-(s_1+s_2+s_3)p'^2 - \frac{k_1^2s_2s_3+k_3^2s_1s_2+k_2^2s_1s_3}{(s_1+s_2+s_3)}.
\]
So we can do the momentum integral (removing prime on $p$):
\[
\int \Dp ~e^{-(s_1+s_2+s_3)p^2 - \frac{k_1^2s_2s_3+k_3^2s_1s_2+k_2^2s_1s_3}{(s_1+s_2+s_3)}}(p-\frac{k_1s_3-k_3s_1}{s_1+s_2+s_3})^\mu (p-\frac{k_1s_3-k_3s_1}{s_1+s_2+s_3})^\nu
\]
The $p^\mu p^\nu$ term gives $\dd^{\mu\nu}$ which vanishes due to tracelessness of $\chi_{\mu\nu}$. Similarly terms involving $k_3^\mu$ or $k_3^\nu$ vanish due to transversality of $\chi_{\mu\nu}$. We are left with
\[
\frac{k_1^\mu k_1^\nu} {(4\pi)^\Dt}\frac{s_3^2}{(s_1+s_2+s_3)^{\Dt+2}}e^{-\frac{k_1^2s_2s_3+k_3^2s_1s_2+k_2^2s_1s_3}{(s_1+s_2+s_3)}}.
\]
So the integral is
\be \label{17k}
I=\frac{k_1^\mu k_1^\nu}{(4\pi)^{\frac{D}{2}}\Gamma(a_1)\Gamma(a_2)\Gamma(a_3)}\int ds_1 ds_2 ds_3 \frac{s_1^{a_1-1}s_2^{a_2-1}s_3^{a_3+ 1}}{(s_1+s_2+s_3)^{\frac{D}{2}+2}}
e^{-\frac{k_1^2s_2s_3+k_3^2s_1s_2+k_2^2s_1s_3}{(s_1+s_2+s_3)}}.
\ee
Change of variables :
\[
s_1=\alpha_1 t,~~s_2=\alpha_2 t,~~s_3=\alpha_3 t,~~~~s_1+s_2+s_3=t,~~~\alpha_1+\alpha_2+\alpha_3=1
\]
We have:
\[ 
\int ds_1ds_2ds_3~=\int dt ~t^2 d\alpha_1d\alpha_2 d\alpha_3 \delta(1-\alpha_1-\alpha_2-\alpha_3).
\]
So
\begin{align}
I= &\frac{1}{(4\pi)^{\frac{D}{2}}\Gamma(a_1)\Gamma(a_2)\Gamma(a_3)}\int dt ~t^2 d\alpha_1d\alpha_2 d\alpha_3\delta(1-\alpha_1-\alpha_2-\alpha_3) \nonumber\\
&(\alpha_1t)^{a_1-1}(\alpha_2t)^{a_2-1}(\alpha_3t)^{a_3+1}t^{-\Dt-2}e^{-k_1^2 \alpha_2\alpha_3t -k_2^2 \alpha_1\alpha_3t -k_3^2 \alpha_2\alpha_1t}.
\end{align}
Further change of variables: 
\be   \label{cov2}
\alpha_1 \alpha_2t=\beta_3,~~\alpha_1 \alpha_3t=\beta_2,~~\alpha_3 \alpha_2t=\beta_1.
\ee
Then
\[
\beta_1\beta_2\beta_3 = (\alpha_1\alpha_2\alpha_3)^2t^3 \implies \frac{(\beta_1\beta_2\beta_3)^\hf}{\beta_3}=\alpha_3 t^\hf.
\]
Thus
\[
(\beta_1\beta_2\beta_3)^\hf(\frac{1}{\beta_1}+\frac{1}{\beta_2}+\frac{1}{\beta_3})=t^\hf
\]
\[ \implies
\frac{(\beta_2\beta_3+\beta_1\beta_3+\beta_2\beta_1)^2}{(\beta_1\beta_2\beta_3)}\equiv \frac{J^2}{\beta_1\beta_2\beta_3}=t.
\]
We have defined
\be   \label{J}
J=(\beta_2\beta_3+\beta_1\beta_3+\beta_2\beta_1).
\ee
Also 
\[
\alpha_3 = \frac{\beta_1\beta_2\beta_3}{J\beta_3} ,~~~\alpha_2=....
\]

So 
\[
\alpha_1^{a_1-1}\alpha_2^{a_2-1}\alpha_3^{a_3-1}= \frac{(\beta_1\beta_2\beta_3)^{a_1+a_2+a_3-3}}{J^{a_1+a_2+a_3-3}}\beta_1^{1-a_1}\beta_2^{1-a_2}\beta_3^{1-a_3},
\]
The change of variables is
\[
\alpha_1 \alpha_2t=\beta_3,~~\alpha_1 (1-\alpha_2 -\alpha_1)t=\beta_2,~~(1-\alpha_2-\alpha_1) \alpha_2t=\beta_1,
\]
Jacobian:
\begin{equation}
J=\det\begin{pmatrix}
\frac{\p \beta_3}{\p \alpha_1} &\frac{\p \beta_3}{\p \alpha_2}&\frac{\p \beta_3}{\p t}\\[6pt]
\frac{\p \beta_2}{\p \alpha_1} &\frac{\p \beta_2}{\p \alpha_2}& \frac{\p \beta_2}{\p t}\\[6pt]
\frac{\p \beta_1}{\p \alpha_1} &\frac{\p \beta_1}{\p \alpha_2}& \frac{\p \beta_1}{\p t}
\end{pmatrix},
\end{equation}
\begin{equation}
J=\det\begin{pmatrix}
\alpha_2 t& \alpha_1t&\alpha_1\alpha_2\\
(1-2\alpha_1 - \alpha_2)t&\alpha_1t&\alpha_1(1-\alpha_1-\alpha_2)\\
-\alpha_2t&(1-\alpha_1-2\alpha_2)t &\alpha_2(1-\alpha_1-\alpha_2)
\end{pmatrix},
\end{equation}
\be
J=\alpha_1\alpha_2 \alpha_3 t^2.
\ee
Putting all this together
\begin{align}
I=\frac{1}{(4\pi)^{\frac{D}{2}}\Gamma(a_1)\Gamma(a_2)\Gamma(a_3)}\int \frac{d\beta_1d\beta_2d\beta_3}{J} t^{a_1+a_2+a_3-\frac{D}{2}-1} \alpha_1^{a_1-1} \alpha_2^{a_2-1} \alpha_3^{a_3+1}\nonumber\\
e^{-k_1^2 \alpha_2\alpha_3t -k_2^2 \alpha_1\alpha_3t -k_3^2 \alpha_2\alpha_1t}\nonumber\\
=\frac{1}{(4\pi)^{\frac{D}{2}}\Gamma(a_1)\Gamma(a_2)\Gamma(a_3)}\int \frac{d\beta_1d\beta_2d\beta_3}{J}\frac{J^{2(a_1+a_2+a_3-\frac{D}{2}-1)}}{J^{a_1+a_2+a_3-1}} \frac{\beta_1^{a_2+a_3}\beta_2^{a_1+a_3}\beta_3^{a_1+a_2-2}}{(\beta_1\beta_2\beta_3)^{a_1+a_2+a_3-1-\Dt}}\nonumber\\
e^{-k_1^2 \beta_1 -k_2^2 \beta_2 -k_3^2 \beta_3}\nonumber\\
I=\frac{1}{(4\pi)^{\frac{D}{2}}\Gamma(a_1)\Gamma(a_2)\Gamma(a_3)}\int d\beta_1d\beta_2d\beta_3 J^{a_1+a_2+a_3-2-D} \beta_1^{\frac{D}{2}+1-a_1}\beta_2^{\frac{D}{2}+1-a_2}\beta_3^{\frac{D}{2}-1-a_3}\nonumber\\
e^{-k_1^2 \beta_1 -k_2^2 \beta_2 -k_3^2 \beta_3}.
\label{I}
\end{align}
Now rewrite
\[
J^{a_1+a_2+a_3-2-D}=(\frac{J}{\beta_1\beta_2\beta_3})^{a_1+a_2+a_3-2-D}(\beta_1\beta_2\beta_3)^{a_1+a_2+a_3-2-D}.
\]
\begin{align*}
I=\frac{1}{(4\pi)^{\frac{D}{2}}\Gamma(a_1)\Gamma(a_2)\Gamma(a_3)}\int d\beta_1d\beta_2d\beta_3(\frac{J}{\beta_1\beta_2\beta_3})^{a_1+a_2+a_3-2-D}\\
\beta_1^{a_2+a_3-1-\Dt}\beta_2^{a_1+a_3-1-\Dt}
\beta_3^{a_1+a_2-3-\Dt}.
\end{align*}
Now use
\begin{align*}
\Big(\frac{J}{\beta_1\beta_2\beta_3}\Big)^{a_1+a_2+a_3-2-D}
=\frac{1}{\Gamma(D+2-a_1-a_2-a_3)}\int_0^\infty dx ~e^{-x(\frac{J}{\beta_1\beta_2\beta_3})}
x^{D+2-a_1-a_2-a_3-1}\\
=
\frac{1}{\Gamma(D+2-a_1-a_2-a_3)}\int_0^\infty dx ~e^{-x(\frac{1}{\beta_1}+\frac{1}{\beta_2}+\frac{1}{\beta_3})}
x^{D+2-a_1-a_2-a_3-1}.
\end{align*}
In interest of getting the appropriate result from the formula \eqref{besselIntegralFormula} below, we rescale $x$.
\begin{equation*}
\Big(\frac{J}{\beta_1\beta_2\beta_3}\Big)^{a_1+a_2+a_3-2-D}=\frac{2^{-2(D+2-a_1-a_2-a_3)}}{\Gamma(D+2-a_1-a_2-a_3)}\int_0^\infty dx ~e^{-\frac x4(\frac{1}{\beta_1}+\frac{1}{\beta_2}+\frac{1}{\beta_3})}
x^{D+2-a_1-a_2-a_3-1}
\end{equation*}
Here $\beta \to \infty$ is the IR region and is cutoff by $k_i^2$. $\beta\to 0$ is the UV end and we can cut it off by cutoffing off the lower end of the $x$ integral by $\frac{1}{\lm^2}$.
Thus
\[
I=\frac{2^{-2(D+2-a_1-a_2-a_3)}}{\Gamma(D+2-a_1-a_2-a_3)}\frac{1}{(4\pi)^{\frac{D}{2}}\Gamma(a_1)\Gamma(a_2)\Gamma(a_3)}\int_0^\infty dx ~x^{D+2-a_1-a_2-a_3-1}
\]
\[ 
\int d\beta_1d\beta_2d\beta_3~\beta_1^{\overbrace{a_2+a_3-\Dt}^{\nu_1}-1}\beta_2^{\overbrace{a_1+a_3-\Dt}^{\nu_2}-1}
\beta_3^{\overbrace{a_1+a_2-2-\Dt}^{-\nu_3}-1}
\]
\be
e^{-k_1^2 \beta_1 -k_2^2 \beta_2 -k_3^2 \beta_3}e^{-\frac x4(\frac{1}{\beta_1}+\frac{1}{\beta_2}+\frac{1}{\beta_3})}
\ee

Now use\cite{Gradshtein}
\begin{equation}
\int_0^\infty
d\beta ~\beta^{\nu-1} e^{-k^2\beta-\frac{x}{4\beta}}=
2^{1-\nu}\Big(\frac{k}{ \sqrt x}\Big)^{-\nu}K_\nu(k \sqrt x).
\label{besselIntegralFormula}
\end{equation}
Therefore
\begin{align}
I=&\frac{2^{-2(D+2-a_1-a_2-a_3)}}{\Gamma(D+2-a_1-a_2-a_3)}\frac{1}{(4\pi)^{\frac{D}{2}}\Gamma(a_1)\Gamma(a_2)\Gamma(a_3)}\int_{\frac{1}{\lm^2}}^\infty dx ~x^{D+2-a_1-a_2-a_3-1}\nonumber\\
&2^{3-\nu_1-\nu_2-\nu_3}\bigg(\frac{k_1}{ \sqrt x}\bigg)^{-\nu_1}K_{\nu_1}(k_1 \sqrt x)~
\bigg(\frac{k_2}{ \sqrt x}\bigg)^{-\nu_2}K_{\nu_2}(k_2 \sqrt x)~
\bigg(\frac{k_3}{ \sqrt x}\bigg)^{\nu_3}K_{\nu_3}(k_3 \sqrt x),
\end{align}
where
\[
\nu_1=a_2+a_3-\Dt,~~~\nu_2=a_1+a_3-\Dt,~~~
-\nu_3=a_1+a_2-\Dt-2.
\]
For the case at hand $a_1=a_2=a_3=1$.
So
\be   \label{nu} 
\nu_1=2-\Dt,~~\nu_2=2-\Dt,~~\nu_3=\Dt.
\ee

Thus we see that
\begin{equation}
\frac{\mathrm dI}{\mathrm dt}=\Lambda \frac{\mathrm d I}{\mathrm d \Lambda}= \sqrt{2}\frac{1}{\Lambda^{2D-2}}(k_1\Lambda)^{-\nu_1}K_{\nu_1}\Big(\frac{k_1}{\Lambda}\Big)~
(k_2\Lambda)^{-\nu_2}K_{\nu_2}\Big(\frac{k_2}{\Lambda}\Big)~
(k_3\Lambda)^{\nu_3}K_{\nu_3}\Big(\frac{k_3}{\Lambda}\Big).
\end{equation}

\end{appendices}

\newpage

\end{document}